\documentclass[acmsmall]{acmart}

\AtBeginDocument{%
  }

\setcopyright{none}
\copyrightyear{2024}
\acmYear{2024}
\acmDOI{XXXXXXX.XXXXXXX}

\renewcommand\footnotetextcopyrightpermission[1]{} 

\acmPrice{15.00}
\acmISBN{978-1-4503-XXXX-X/18/06}

\usepackage{enumitem}
\usepackage{xcolor}
\usepackage[many]{tcolorbox}
\usepackage{multirow}

\usepackage[font=small,skip=3pt]{caption}
\newcommand{\distance}{8pt}
\usepackage{balance,wrapfig}

\usepackage{listings}
\definecolor{codegreen}{rgb}{0,0.6,0}
\definecolor{codegray}{rgb}{0.5,0.5,0.5}
\definecolor{codepurple}{rgb}{0.58,0,0.82}
\definecolor{backcolour}{rgb}{0.95,0.95,0.92}

\lstdefinestyle{modifiedpython}{
    basicstyle=\ttfamily\footnotesize,
    breaklines=true,
    captionpos=b,
    xleftmargin=2em,
    xrightmargin=2em,
    columns=fullflexible,
    frame=lines,
    showstringspaces=false,
    commentstyle=\color{gray},
    keywordstyle=\color{blue},
    stringstyle=\color{orange},
    keepspaces=true,
}
\newcommand{\swj}[1]{\textcolor{red}{#1}}
\newcommand{\pr}[2]{\href{#1}{\textcolor{blue}{PR\#{#2}}}}
\newcommand{\issue}[2]{\href{#1}{\textcolor{blue}{Issue\#{#2}}}}

\newcommand{\lstbg}[3][0pt]{{\fboxsep#1\colorbox{#2}{\strut #3}}}
\definecolor{codegreen}{rgb}{0,0.6,0}
\lstdefinelanguage{diff}{
	frame=lines,
	basicstyle=\ttfamily\scriptsize,
	morecomment=[f][\color{red}]{---}, 
	morecomment=[f][\color{codegreen}]{+++},
	morecomment=[f][\lstbg{red!20}]{-\ },
	morecomment=[f][\lstbg{green!20}]{+\ },
	morecomment=[f][\color{blue}]{@@},
}

\begin{document}

\title{A Comprehensive Empirical Study of Bugs in Open-Source Federated Learning Frameworks}

\author{Weijie Shao}
\affiliation{%
  \institution{ShanghaiTech University}
    \country{China}
}

\author{Yuyang Gao}
\affiliation{%
  \institution{ShanghaiTech University}
    \country{China}
}

\author{Fu Song}
\authornote{Corresponding author.}
\email{songfu@ios.ac.cn}
\affiliation{%
	\institution{State Key Laboratory of Computer Science, Institute of Software, Chinese Academy of Sciences \& University of Chinese Academy of Sciences}
	\city{Beijing}
	\country{China}
	\postcode{100190}
} 

\author{Sen Chen}
\affiliation{%
  \institution{Tianjin University}
\country{China}
}
\author{Lingling Fan}
\affiliation{%
  \institution{Nankai University}
 \country{China}
}

\author{Jingzhu He}
\affiliation{%
  \institution{ShanghaiTech University}
\country{China}
}



\renewcommand{\shortauthors}{Shao et al.}

\begin{abstract}

Federated learning (FL) is a distributed machine learning (ML) paradigm, 
allowing multiple clients to collaboratively train shared machine learning (ML) models 
without exposing clients' data privacy. It has gained substantial popularity in recent years, especially since the enforcement of data protection laws and regulations in many countries.
To foster the application of FL,
a variety of FL frameworks have been proposed,
allowing non-experts to easily train ML models. 
As a result, understanding bugs in FL frameworks is critical for 
facilitating the development of better FL frameworks and potentially encouraging the development of bug detection, 
localization and repair tools.
Thus, we conduct the first empirical study to comprehensively collect,
taxonomize, and characterize bugs in FL frameworks.
Specifically, we manually collect and classify 
1,119 bugs from all the 676 closed issues and 514
merged pull requests in 17 popular and representative open-source FL frameworks on GitHub.  
We propose a classification of those bugs 
into 12 bug symptoms, 12 root causes, and 18 fix patterns.
We also study their correlations and distributions on 23 functionalities. 
We identify nine major findings from our study,
discuss their implications and 
future research directions based on our findings.
\end{abstract}



\maketitle
\section{Introduction}

To break the barriers between isolated data
and protect data privacy,
federated learning (FL) was proposed by Google~\cite{KonecnyMR15} as a distributed machine learning (ML) paradigm for collaboratively training shared ML models with the private data kept in the clients locally~\cite{LiSM20}.
With the increasing concerns on
data privacy,
various data protection laws and regulations have been enacted,
e.g., European Union's General Data Protection Regulation (GDPR)~\cite{european_commission_regulation_2016} and China’s Data Security Law~\cite{china_data_security_law}.
As a result, different FL privacy protection techniques have been proposed (cf.~\cite{YinZH21} for a survey) 
and a number of FL-based systems have been developed and deployed~\cite{YangLCT19,abs-2104-10501}.
To facilitate the application of FL, a variety of open-source FL frameworks have been developed, e.g., 
TensorFlow Federated (TFF)~\cite{BonawitzEGHIIKK19}, and NVIDIA NVFlare~\cite{RothCWY0HKHZL0023}.
%
While the security of FL-based systems has been studied (e.g.,~\cite{BonawitzIKMMPRS17, MohasselZ17, GeyerKN17, FungYB18, PhongAHWM18, FangCJG20, XuLL0L20, ZhangLXWYL20, WengWZLZL21}),
bugs in FL frameworks have not been systematically investigated yet.
A comprehensive understanding of such bugs is essential for improving the quality of FL frameworks. 
Previous works have studied bug characteristics in a variety of domains including numerical software libraries~\cite{FrancoGR17},
ML/DL libraries~\cite{ThungWLJ12,IslamNPR19,ZhangCCXZ18,JiaZWHL21}, 
concurrency bugs~\cite{LuPSZ08,Leesatapornwongsa16},
performance bugs~\cite{JinSSSL12,NiariC020}, 
error-handling bugs~\cite{CachoBAPGCSCFG14,CoelhoAGD15},
and autonomous vehicle bugs~\cite{GarciaF0AXC20}.
These studies provide insights and guidelines for improving  system quality, however, none of existing studies have focused on bugs in FL frameworks.


%

\smallskip
\noindent
{\bf Research questions.} 
This work presents the first comprehensive study of bugs in FL frameworks, addressing the following research questions:

\begin{enumerate}[label=\textbf{RQ\arabic*.},itemindent=*,leftmargin=*,itemsep=0pt]
\item {\it What are the symptoms (effects) that bugs exhibit and the distribution of different symptoms in FL frameworks?} 
\item {\it What are the root causes of bugs that reflect mistakes developers make, the distribution of root causes, and the correlation to symptoms?}
\item {\it What are the fix patterns used by developers to fix bugs, the distribution of fix patterns, and correlation to root causes?} 
\item {\it What are the distributions of symptoms, root causes, and fix patterns 
across functionalities (e.g., data collection, data preprocessing)?} 
\end{enumerate}


Answering these research questions 
provides kinds of information that 
is useful for understanding the consequences of bugs,
developing better FL frameworks, and creating bug detection, localization and repairing tools~\cite{ZhangCCXZ18,ZhangGMLK19,ShenM0TCC21,YangHXF22,QuanGXCLL22,GarciaF0AXC20}.

To answer these research questions, 
we select 17 popular and representative open-source FL frameworks
on GitHub 
(FATE~\cite{LiuFCXY21}, 
PySyft~\cite{RyffelTDWMRP18}, 
FederatedScope~\cite{federatedscope},
NVFlare~\cite{RothCWY0HKHZL0023}, 
OpenFL~\cite{reina2021openfl}, 
Plato~\cite{plato-iwqos22}, 
FEDn~\cite{EkmefjordAAASST22}, 
Flower~\cite{beutel2020flower}, 
TensorFlow Federated (TFF)~\cite{BonawitzEGHIIKK19}, 
FedScale~\cite{fedscale-icml22}, 
PrimiHub~\cite{PrimiHub},
APPFL~\cite{appfl-ipdps22},
flame~\cite{flame},
FedML~\cite{he2020fedml},
FedTree~\cite{FedTree},
RayFed~\cite{RayFed},
and FedSim~\cite{FedSim}),
ranging from large popular industrial products to small academic research ones.
We collected all the 676 closed issues and 514
merged pull requests in these frameworks, before February~22,~2023. 

\smallskip
\noindent
{\bf Challenges.}
However, there remain several challenges that are time-consuming and labor-intensive to be addressed. The two major challenges are:

\begin{itemize}[leftmargin=*]
    \item {\bf Challenge 1: Disorganized Pull Requests \& Ambiguous Bug Descriptions}.
Almost all the FL frameworks either do not enforce strict bug report formats or the provided bug report templates are not used consistently, resulting in disorganized pull requests and ambiguous bug descriptions. As a result, multiple irrelevant bugs may be fixed in one disorganized pull request and a single bug may be fixed in multiple disorganized pull requests, increasing the difficulty of data collection and analysis. Ambiguous bug descriptions in the pull requests, issues, or commit messages also significantly increase the difficulty of data collection, classification and labeling.
   \item {\bf Challenge 2: Heterogeneous Architecture}.
The development of the FL frameworks is a relatively emerging domain. 
Different development teams have their own goals and ideas,
thus adopt different architecture designs for extensibility~\cite{RothCWY0HKHZL0023}, scalability~\cite{BonawitzEGHIIKK19}, and user-friendliness~\cite{beutel2020flower}
with different techniques (e.g., secure multi-party computation, differential privacy
and homomorphic encryption) to provide meaningful privacy guarantees~\cite{YangLCT19}.
In particular, the architectures of the auxiliary functionalities vary greatly
and bugs in the auxiliary functionalities account for a considerable percentage. 
Thus, the heterogeneous architectures among FL frameworks pose a great challenge for the construction of taxonomies.
\end{itemize}

To address Challenge 1, 
we identify the commits with bug-fixing keywords and record fixing descriptions to track commits that fix the same bug in disorganized pull requests. For ambiguous bug descriptions, we try our best to understand the code context and infer the possible label. If the code context of a bug is simple, we try to write a similar code snippet to reproduce the possible symptoms. After identifying all the bugs in these issues and requests, we obtained 1,112 bugs on which we 
conducted the empirical study.

To address Challenge 2, 
we adopt a widely used open-coding scheme~\cite{Seaman99} to construct new taxonomies with the cooperation of two authors. Sometimes when a new consensus that comparatively differs from the previous one is reached, we go through the dataset from the beginning again to alter existing intermediate classification and labeling results.

\smallskip
\noindent
{\bf Main contributions.}
In summary, this paper makes the following main contributions:
\begin{itemize}[leftmargin=*]
    \item We conduct the first comprehensive investigation of bugs in FL frameworks through an empirical study of
    1,119 bugs from 17 popular and representative open-source FL frameworks. 
    \item We classify all the 1,119 bugs and provide a systematic classification of these bugs with 12 symptoms, 12 root causes, 18 fix patterns.
%
    \item We analyze and present the distributions and correlations of symptoms, root causes, and fix patterns, and their differences across 23 functionalities. 
%
    \item We summarize 9 findings from our study, 
    discuss and suggest future research directions for practitioners and researchers based on our findings. 
%
\end{itemize}
 
\noindent{\bf Findings-at-a-glance.}
Our findings are interesting and informative.
For instance, \textit{Functional Error} (544/1,119) and \textit{Crash} (341/1,119) are the
two most frequent symptoms, indicating that 
there is a lack of systematic testing tools for  
FL frameworks.
Some bugs (7/1,119) will bring the privacy leakage problem that violates one of the essential properties of federated learning, thus calling for data privacy analysis tools.
Compared with in DL frameworks, {\it Privacy Leak},
{\it Security Vulnerability}, and {\it Flaky Test} are unique symptoms in FL frameworks. 
\textit{Logic Error} (418/1,119) is the most frequent root cause of bugs whose fixes are often very involved, but a large number of bugs are caused by simple assignment problems (318/1,119) due to careless programming. 
Bugs caused by typos (61/1,119) mostly fail to pass syntax checking, indicating that developers may push source code
into repositories without a local compilation.
The two most frequent symptoms \textit{Functional Error} and \textit{Crash} are positively correlated with diverse root causes
while the two most frequent root causes \textit{Logic Error} and \textit{Assignment Problem} also result in a broad range of symptoms, posing challenges to bug detection and localization.
Most bugs (736/1,119) occur in auxiliary functionalities such as utilities and examples instead of the core functionalities, indicating that developers should pay more attention to the auxiliary functionalities.



\smallskip
 \noindent{\bf Outline}.
  Section~\ref{sec:methodology} describes the process of data collection, classification, and labeling.
  Section~\ref{sec:results} reports our results and findings.
  Section~\ref{sec:threats2validity} discusses the threats to validity.
We discuss related work in Section~\ref{sec:related_work} 
and conclude this paper in Section~\ref{sec:conclusion}.

\section{Methodology} \label{sec:methodology}

\subsection{Selection of FL Frameworks}
To comprehensively investigate the characterizations of bugs in FL frameworks, we should collect a large number of bugs from popular and representative FL frameworks while keeping diversity at the same time.
Following prior work~\cite{GarciaF0AXC20,ShenM0TCC21,QuanGXCLL22}, we use GitHub search to collect open-source FL frameworks. Initially, we apply the following criteria to select repositories: 
(1) owned and maintained by a GitHub organization account;
(2) not experiments of some papers;
(3) not a demo, tutorial or benchmark; and
(4) not archived by the owner.
As a result, 45 frameworks are selected.

\begin{wraptable}{R}{6.5cm}
\setlength{\tabcolsep}{2pt}
\caption{Collected Data of Open-Source FL Frameworks, ordered by the number of bugs.}
\label{tab:frameworks-overview}
\scalebox{.7}{ 
\begin{tabular}{l|c|cc|cc|c}
\hline
\multirow{2}{*}{\textbf{Framework}}  & \multirow{2}{*}{\textbf{Star}} & \multicolumn{2}{c|}{\textbf{Original Data}} & \multicolumn{2}{c|}{\textbf{Final Data}} & \multirow{2}{*}{\textbf{\#Bug}} \\ \cline{3-6}
                                     &                                 & \textbf{\#Issue}       & \textbf{\#PR}      & \textbf{\#Issue}     & \textbf{\#PR}     &                                  \\ \hline
FATE~\cite{LiuFCXY21}                & 5.1k                            & 115                    & 307                & 81                   & 295               & 733                              \\
PySyft~\cite{RyffelTDWMRP18}         & 8.9k                            & 338                    & 55                 & 85                   & 113               & 132                              \\
FederatedScope~\cite{federatedscope} & 981                             & 31                     & 65                 & 37                   & 71                & 81                               \\
NVFlare~\cite{RothCWY0HKHZL0023}     & 409                             & 34                     & 38                 & 28                   & 55                & 57                               \\
OpenFL~\cite{reina2021openfl}        & 564                             & 20                     & 20                 & 15                   & 29                & 30                               \\
Plato~\cite{plato-iwqos22}           & 237                             & 21                     & 0                  & 17                   & 18                & 20                               \\
FEDn~\cite{EkmefjordAAASST22}        & 96                              & 17                     & 12                 & 14                   & 18                & 18                               \\
Flower~\cite{beutel2020flower}       & 2.8k                            & 21                     & 10                 & 7                    & 20                & 16                               \\
TFF~\cite{BonawitzEGHIIKK19}         & 2.1k                            & 63                     & 0                  & 14                   & 14                & 14                               \\
FedScale~\cite{fedscale-icml22}      & 312                             & 4                      & 0                  & 4                    & 3                 & 5                                \\
PrimiHub~\cite{PrimiHub}             & 970                             & 2                      & 4                  & 1                    & 5                 & 4                                \\
APPFL~\cite{appfl-ipdps22}           & 34                              & 3                      & 0                  & 3                    & 2                 & 3                                \\
flame~\cite{flame}                   & 34                              & 2                      & 0                  & 2                    & 2                 & 2                                \\
FedML~\cite{he2020fedml}             & 3.1k                            & 2                      & 2                  & 1                    & 2                 & 1                                \\ 
FedTree~\cite{FedTree}               & 115                             & 1                      & 0                  & 1                    & 1                 & 1                                \\ 
RayFed~\cite{RayFed}                 & 57                              & 1                      & 0                  & 1                    & 1                 & 1                                \\ 
FedSim~\cite{FedSim}                 & 7                               & 1                      & 1                  & 0                    & 1                 & 1                                \\ 
\hline
\textbf{Total}                       & \textbf{-}                      & \textbf{676}           & \textbf{514}       & \textbf{312}         & \textbf{650}      & \textbf{1,119}                   \\ 
\hline
\end{tabular}

}
\end{wraptable}

Next, we filter out 13 inactive FL frameworks whose default branches have no commits in the past 180 days.
Among the 32 remaining FL Frameworks: 
FLSim~\cite{FLSim},
Fedlab~\cite{FedLab},
FedSim~\cite{FedSim},
Galaxy Federated Learning Framework~\cite{GFL},
PaddleFL~\cite{PaddleFL},
XFL~\cite{XFL},
FedJAX~\cite{FedJAX},
Fedlearner~\cite{Fedlearner},
fl-simulation~\cite{fl-simulation},
iFLearner~\cite{iFLearner},
FLUTE~\cite{FLUTE},
SecretFlow~\cite{SecretFlow},
and Substra~\cite{Substra}
are excluded due to the lack of merged pull requests and issues that are tagged with ``bug''; IBM federated learning~\cite{ibm-federated-learning} is excluded because the repository contains a Python Wheel file only without source code; and Mindspore~\cite{mindspore23} is excluded  due to the lack of independent FL framework repository.
Finally, the remaining 17 open-source FL frameworks 
are selected and used
in our study as shown in Table~\ref{tab:frameworks-overview}, ranging from large popular industrial products to small academic research ones.

\subsection{Collection of Bug Data}
From the 17 frameworks, we collect 
all the 676 closed issues and 
all the 514 merged pull requests (written as PR in Table~\ref{tab:frameworks-overview}) that are tagged with ``bug'',
before February~22,~2023.
We focus on merged pull requests because they
contain code changes, discussions, links to
related issues, and other information 
that are vital for a comprehensive understanding of bugs and their fixes.

In the initial review of pull requests and issues,
we found that 
\begin{itemize}[leftmargin=*]
    \item 396 issues with ``bug'' tag are not linked to by any merged pull requests 
    (e.g., \issue{https://github.com/OpenMined/PySyft/issues/118}{118} in PySyft), thus do not have any code changes and discussions;
    \item 32 issues without ``bug'' tag actually
    report bugs and are linked to by pull requests with ``bug'' tag  
    (e.g., \issue{https://github.com/alibaba/FederatedScope/issues/91}{91} in FederatedScope), 
    thus the relevant bugs and their fixes should be identified;
    \item 136 merged pull requests without ``bug'' tag fix bugs 
    and link to related issues with ``bug'' tag 
    (e.g., \pr{https://github.com/OpenMined/PySyft/pull/2560}{2560} in PySyft), 
    thus the relevant bugs and their fixes should also be identified;
    \item 133 merged pull requests individually fixes multiple irrelevant bugs
    (e.g., \pr{https://github.com/FederatedAI/FATE/pull/694}{694} in FATE), 
    thus multiple individual bugs should be identified.
\end{itemize}


After an in-depth analysis, 
we finally identify all the 1,119 bugs 
that come from the 312 closed issues and 650 merged pull requests, as shown in Table~\ref{tab:frameworks-overview}.

\subsection{Classification and Labeling Process of Bug Data}
To characterize bugs with reduced subjectivity bias, 
%
two authors cooperate to complete the taxonomy and labeling, following an open-coding scheme~\cite{Seaman99} that has been widely used in
establishing classification taxonomies~\cite{GarciaF0AXC20, ShenM0TCC21, QuanGXCLL22}.
Due to the large number of bugs, we conduct a two-phase classification
and labeling process.

In the first phase, a random sampling method is adopted to select 30\% of bugs ($1,119\times 30\%\approx336$) for pilot taxonomy 
and the two authors engage in extensive discussions to refine and adjust the taxonomy whenever some bugs are not definitively assigned to a specific category.

In the second phase, 
the established pilot taxonomies are utilized by the two authors to taxonomize the remaining bugs through multiple rounds of assessment. 
During each round, following prior work~\cite{IslamNPR19, ShenM0TCC21, QuanGXCLL22}, a coefficient measuring agreement is calculated
between the classification results of the two authors. 
Due to the diversity and larger number of
bugs involved in this phase, initially, the disagreement of results between the two authors leads to a Cohen's Kappa coefficient~\cite{cohen1960coefficient} of approximately 53\%.
To address this issue, the authors revisit the taxonomies by separating some large categories into small ones and adjusting ambiguous categories. Based on the resulting taxonomies, several additional rounds of discussion and adjustment are repeated. As a result of this rigorous process, the Cohen's Kappa coefficient substantially improves from 53\% to over 92\% in the rounds formally recorded within the final dataset. In these rounds, all the bugs are confidently and unequivocally categorized without dispute.

\subsection{Correlation Metric} 
To investigate the correlation between symptoms, root causes and fix patterns,
we adopt a statistical metric, called
\textit{lift}~\cite{HanKP2011}. 
The \textit{lift} metric has been used in several works~\cite{TanLLWZZ14,JiaZWHL21,DuSLA23}.
The lift value $\textit{lift}(A, B)$ between categories $A$ and $B$ is 
   $$\frac{P(A \cap B)}{P(A)\times P(B)}$$
where $P(A)$ and $P(B)$ denote the probabilities that a bug belongs to the categories $A$ and $B$, respectively; and $P(A \cap B)$ denotes the probability that a bug belongs to both $A$ and $B$.

From the lift value $\textit{lift}(A, B)$, we can analyze the correlation
between categories $A$ and $B$.
Specifically, if $\textit{lift}(A, B)=1$, then categories $A$ and $B$
have no correlation;
if $\textit{lift}(A, B) > 1$, categories $A$ and $B$ are positively correlated, i.e.,
a bug of category $A$ is likely to be of category $B$; 
if $\textit{lift}(A, B) < 1$, categories  $A$ and $B$ are negatively correlated, i.e.,
a bug of category $A$ is unlikely to be of category $B$.
 %
Consider $100$ bugs, where $60$ bugs result in {\it crash} and $40$ bugs are caused by {\it logic error}. If the number of crash bugs caused by {\it logic error} is $30$, then $P(crash)=60/100$, $P(logic~ error)=40/100$ and $P(crash\cap logic~ error)=30/100$. 
$\textit{lift}(crash, logic ~error)=(30/100)/((60/100)\times(40/100))=1.25$, meaning
that crash bugs are more likely caused by {\it logic error}.


\section{Results} \label{sec:results}



This section provides an in-depth analysis of the outcomes from our empirical study and 
answers the research questions. 
Here we mainly present the primary categories of the taxonomies while the detailed sub-categories are given in our dataset~~\cite{FLBugdataset}.
We remark that the cumulative probabilities of all categories within root causes, fix patterns, or functionalities may be greater than 100\% due to that one single bug can contain more than one root cause, fix pattern, or functionality.



\subsection{RQ1: Symptoms}
We identified the following 12 main symptoms of the 1,119 bugs (including the effects of bugs).
\begin{figure}[t]
    \centering
\begin{lstlisting}[
        language=diff,
        label=lst:BldErr, 
        caption={Build Error (\issue{https://github.com/FederatedAI/FATE/issues/3910}{3910} in FATE)}
    ]
  # python/fate_arch/common/file_utils.py
+ READTHEDOC = os.getenv("READTHEDOC")
  def get_project_base_directory(*args):
      ...
+     global READTHEDOC
      if PROJECT_BASE is None:
      ...
+         if READTHEDOC is None:
+             PROJECT_BASE = os.path.abspath(os.path.join(PROJECT_BASE, os.pardir))
      ...
      return PROJECT_BASE
\end{lstlisting}
    \vspace{-2mm}
\end{figure}


\begin{enumerate}[leftmargin=*]
    \item {\bf Build Error (BldErr)} occurs when the preprocessing, compilation, or linking fails. 
    For example, 
    \issue{https://github.com/FederatedAI/FATE/issues/3910}{3910} reports a \texttt{mkdocs} build failure in FATE, which is shown in Code Snippet~\ref{lst:BldErr}.
%
%
    \item {\bf Crash} 
    terminates a running program improperly 
    often ending with error messages but sometimes silently. This category consists of many sub-categories according to the specific unexpected exceptions. 
    \item {\bf Flaky Test (FlkyTst)} are
    inconsistent results when no changes are made to the program under test and a test
    case. It is only reported in PySyft, where the \verb|test_serde_simplify| test case frequently fails 
    (sometimes but not always) in \issue{https://github.com/OpenMined/PySyft/issues/3359}{3359}.
%
    \item {\bf Functional Error (FuncErr)}
    occurs when the (intermediate) results are inconsistent with the expected ones. 
    The two most frequent sub-categories are {Unexpected Return Value} and {Unexpected Field Value}.
    Note that some bugs with functional errors finally result in crashes,
    thus are categorized into Crash instead of Functional Error. 
%
%
    \item {\bf Hang} happens when the running program never stops.
    For example, \issue{https://github.com/NVIDIA/NVFlare/issues/193}{193}
    reports a bug of NVFlare that {\itshape ``clients don't shut down after end of run''}. 
%
    \item {\bf Installation Failure (InstFl)}
    occurs when the installation of a framework fails.
    For instance, \issue{https://github.com/OpenMined/PySyft/issues/125}{125}
    reports a failure of installing PySyft on MacOS, accompanied by the following error message: {\it ``\texttt{src/gmpy.h}:252:12: fatal error: `\texttt{mpfr.h}' file not found''}.
    %
%
%
    \item {\bf Memory Leak (MemLeak)} happens when developers fail to free some allocated but unused memory. 
    For example, 
    \issue{https://github.com/OpenMined/PySyft/issues/3397}{3397} and \issue{https://github.com/OpenMined/PySyft/issues/3398}{3398},
    respectively, report a bug in the \texttt{relu()} and \texttt{argmax()} methods of PySyft resulting in Memory Leak, which happen {\itshape ``only on the first worker in the worker list passed to the \texttt{encrypt()}.''}
    \item {\bf Performance Degradation (PerfD)} occurs when a system unexpectedly runs slowly or consumes more resources. 
    Besides,
    Poor Accuracy (e.g., 
    \issue{https://github.com/TL-System/plato/issues/146}{146}
    in Plato) and Unexpected Loss Values (e.g., 
    \issue{https://github.com/OpenMined/PySyft/issues/2642}{2642}
    in PySyft) caused by bugs in FL frameworks, are 
    regarded as two sub-categories of Performance Degradation. 
%
    \item {\bf Privacy Leak (PrvLeak)} refers to the incorrect behaviors
    that violate the protection of data privacy.
    Since data privacy is one of the most essential properties of FL-based systems,
    we categorize it as an individual major symptom out of Security Vulnerability.
    For instance, a violation of data privacy protection is reported in
    \issue{https://github.com/FederatedAI/FATE/issues/869}{869} in FATE
    that {\it ``the host sends its \texttt{forwards} and \texttt{loss\_regular} directly without encryption.''}, which is shown in Code Snippet~\ref{lst:PrvLeak}.
    \item {\bf Security Vulnerability (SecVul)}
    refers to the results that  bugs are vulnerable 
    excluding those undermining data privacy directly.
    For example, 
    \issue{https://github.com/OpenMined/PySyft/issues/6965}{6965}
    in PySyft reports that the public argument \texttt{id\_at\_location\_override} of the \texttt{send} method allows adversaries to delete or modify the DataOwner or the uploaded objects.  
    \item {\bf Unknown (Unk)}
    refers to the symptom of bugs that cannot be pinpointed because of unclear descriptions and complicated contexts. 
    \item {\bf Warning (Warn)}
    refers to the behaviors where the running program prints warning messages. 
    For instance,
    in \issue{https://github.com/tensorflow/federated/issues/1333}{1333},
    a Keras model 
    in TFF 
    throws: {\it ``WARNING:tensorflow:Please add \texttt{keras.layers.InputLayer} instead of \texttt{keras\-.Input} to Sequential model.
    \texttt{keras.Input} is intended to be used by Functional model''}.
%
\end{enumerate}

\begin{figure}[t]
    \centering
\begin{lstlisting}[
        language=diff, 
        label=lst:PrvLeak, 
        caption={Privacy Leak (\issue{https://github.com/FederatedAI/FATE/issues/869}{869} in FATE)}
    ]
  # federatedml/optim/gradient/hetero_lr_gradient_and_loss.py
- def compute_loss(self, lr_weights, optimizer, n_iter_, batch_index):
+ def compute_loss(self, lr_weights, optimizer, n_iter_, batch_index, cipher_operator):
      ...
-     self.remote_loss_intermediate(self_wx_square, suffix=current_suffix)
+     en_wx_square = cipher_operator.encrypt(self_wx_square)
+     self.remote_loss_intermediate(en_wx_square, suffix=current_suffix)
      
      loss_regular = optimizer.loss_norm(lr_weights)
-     self.remote_loss_regular(loss_regular, suffix=current_suffix)
+     en_loss_regular = cipher_operator.encrypt(loss_regular)
+     self.remote_loss_regular(en_loss_regular, suffix=current_suffix)
\end{lstlisting}\vspace{-3mm}
\end{figure}

\begin{table}[H]
\centering \vspace{-2mm} \setlength{\tabcolsep}{4pt}
\caption{Distribution of bugs w.r.t. symptoms, ordered by their percentages.} 
\label{tab:symptom-distribution}
\scalebox{.8}{ 
\begin{tabular}{lcc|lcc}
\toprule
{\bf Symptom}                       & {\bf \#Bug} & {\bf Percentage} & {\bf Symptom}                   & {\bf \#Bug} & {\bf Percentage}   \\ 
\midrule
Functional Error (FuncErr)          & 544            & 48.6\%         & Crash                           & 341            & 30.5\%           \\
Unknown (Unk)                       & 99             & 8.8\%          & Build Error (BldErr)            & 36             & 3.2\%            \\
Performance Degradation (PerfD)     & 27             & 2.4\%          & Hang                            & 18             & 1.6\%            \\
Installation Failure  (InstFl)      & 17             & 1.5\%          & Security Vulnerability (SecVul) & 11             & 1.0\%            \\
Privacy Leak (PrvLeak)              & 7              & 0.6\%          & Warning (Warn)                  & 7              & 0.6\%            \\
Flaky Test (FlkyTst)                & 6              & 0.5\%          & Memory Leak (MemLeak)           & 5              & 0.4\%            \\
\midrule
&                                 &                &  {\bf Total}                         & {\bf 1,119}    & {\bf 100.0\%}                   \\ 
\bottomrule
\end{tabular}

}\vspace{-1mm}
\end{table}

Table~\ref{tab:symptom-distribution} shows the statistics of the identified symptoms. 
%
Compared to the symptoms of bugs in DL frameworks, Privacy Leak is one of the symptoms that is specific to the domain of FL frameworks.
There are 7 bugs resulting in privacy leakage: 6 bugs from the industrial FL framework FATE
and 1 bug from the academic FL framework Plato. 
Although such bugs only account for 0.6\% (7/1,119), they violate one of the essential properties of FL, i.e., data privacy.
It is notable that privacy leakage is hard to be caught and 
requires dedicated testing and detection techniques. Thus, it may be more frequent in practice than reported here.

Among the symptoms in our classification, Functional Error 
is the most frequent symptom, accounting for 48.6\% (544/1,119).  
A large number of bugs with vague descriptions of symptoms fall into one of the two most frequent sub-categories of Functional Error, 
i.e., Unexpected Return Value and Unexpected Field Value, accounting for 36.8\% (200/544) and 27.6\% (150/544) of Functional Error bugs, respectively. 
The second and third most frequent symptoms
are Crash and Build Error, 
accounting for 30.5\% (341/1,119) and 3.2\% (36/1,119) of bugs, respectively.  
 
 
Performance degradation is another frequent and notable type of symptom with 27 bugs. 
Besides the poor accuracy in prediction, such bugs in FL frameworks
also consume communication bandwidth and disk storage in training stage due to complicated techniques involved in FL.
For example, a bug reported in 
\issue{https://github.com/tensorflow/federated/issues/896}{896}
in TFF leads to redundant decompression in \texttt{tff.simulation.datasets.emnist.load\_data()}
that significantly affects execution time,
and 
\issue{https://github.com/FederatedAI/FATE/issues/3765}{3765}
in FATE reports that {\it ``\texttt{Stats} object to store is costly''}.

 %
 

\begin{tcolorbox}[size=title]   
    {\bf Finding 1:}
 Some bugs can cause privacy leakage even in
industrial FL frameworks. 
    {\it Functional Error} and {\it Crash} are the two most frequent symptoms, 
    covering 48.6\% (544/1,119) and 30.5\% (341/1,119) of bugs, respectively.
   8.8\% (99/1,119) of bugs cannot be categorized (i.e., unknown) due to unclear descriptions in issue reports and merged pull requests. 

\end{tcolorbox} 

\smallskip
\noindent
{\bf Implication:} This finding indicates
that many bugs (8.8\%) are reported without clear descriptions of symptoms, which is a heavy burden for developers. Designing and enforcing strict bug report formats could help developers to deal with such bugs. 
82.3\% (48.6\%+30.5\%+3.2\%) of bugs result in functional error, crash and build error, we suggest developers
conducting sufficient testing before releasing a version.
Some bugs lead to privacy leakage and performance degradation, which are non-functional properties, thus calling for dedicated analysis techniques and tools.

\smallskip
\noindent
{\bf Comparison with Bug Symptoms in DL Frameworks}.
(1) Privacy Leak, Security Vulnerability, and Flaky Test are unique symptoms of bugs in FL frameworks;
(2) Though Functional Error and Crash are also the two most frequent symptoms in DL frameworks~\cite{JiaZWHL21, YangHXF22, ChenLSQJL23, DuSLA23},
Functional Error is the most frequent one in~\cite{JiaZWHL21,YangHXF22} and Crash is the most frequent one in~\cite{ChenLSQJL23,DuSLA23};
and (3) Build Error is also the third most frequent one in DL frameworks~\cite{JiaZWHL21, ChenLSQJL23}.

\subsection{RQ2: Root Causes and their Correlation to Symptoms}
\subsubsection{Root Causes}
We identified the following 12 root causes of bugs. 

\begin{enumerate}[leftmargin=*]
    \item {\bf API Error (APIErr)} consists of two sub-categories: 
    \textit{API Missing} and \textit{API Misuse}. 
    API Missing leads to bugs when expected APIs, including decorators, are not called or imported, e.g., 
    \pr{https://github.com/OpenMined/PySyft/pull/1315}{1315}
    in PySyft reveals that 
    \texttt{NameError} is caused by a missing import of \texttt{FloatTensor}. 
    API Misuse refers to the calling and importing of incorrect APIs, e.g., 
    \pr{https://github.com/OpenMined/PySyft/pull/3525}{3525} 
    shows that \texttt{initialize\_crypto\_plans} is misused in PySyft. 

    \item {\bf Assignment Problem (AsgProb)} has two sub-categories: \textit{Assignment Error} and \textit{Assignment Missing}.
    Assignment Error causes bugs where variables/parameters
    are assigned incorrectly or redundantly, e.g.,
    \pr{https://github.com/scaleoutsystems/fedn/pull/277}{277} reports that 
    \texttt{NameError} is caused by assigning \texttt{run\_path} instead of \texttt{self.run\_path} in EFDn. 
    Assignment Missing refers to cases where the assignments to variables/parameters are missing, e.g.,
    \issue{https://github.com/alibaba/FederatedScope/issues/221}{221} reports that 
    the assignment \texttt{return\_raw=True} is missing in FederatedScope. 
 %
%
    \item {\bf Computation Error (CompErr)} refers to incorrect 
    numerical computation expressions (e.g., 
    \pr{https://github.com/OpenMined/PySyft/pull/139}{139}
    in PySyft) and incorrect relational computation expressions (e.g., 
    \issue{https://github.com/alibaba/FederatedScope/issues/125}{125}
    in FederatedScope).
    We remark that incorrect relational computation expressions are regarded
    as Condition Error (see below) when they are used as condition expressions.
    \item {\bf Concurrency Error (CncyErr)} includes incorrect handling of multiprocessing or multi-threading. 
    For example, a race condition, 
    {\itshape ``a fed event could be fired after \verb|END_RUN| event''}, is reported in
    \issue{https://github.com/NVIDIA/NVFlare/issues/52}{52} in NVFlare.
%
    \item {\bf Condition Problem (CndProb)} consists of two sub-categories: \textit{Condition Error} and \textit{Condition Missing}.
    The former refers to incorrect and useless conditions (e.g., 
    \issue{https://github.com/OpenMined/PySyft/issues/4616}{4616}
    in PySyft)
    while the latter refers to the lack of additional conditions  (e.g.,
    \issue{https://github.com/OpenMined/PySyft/issues/3435}{3435} reports that
    the condition \texttt{location == destination} is missing 
    in PySyft~\cite{cnd_missing}),
     in condition expressions of loop or branching statements. 
    \item {\bf Documentation Error (DocErr)} refers to incorrect, missing, or disused descriptions/instructions in documentations,  
    confusing or misleading users. 
    For example, the installation failure
    reported in \issue{https://github.com/OpenMined/PySyft/issues/125}{125}
    in PySyft
    results from the lack of instructions 
    for installing prerequisite libraries.
%
    \item {\bf Exception Handling Error (ExcHdErr)} includes
    the lack of exception handlers and incorrect implementations of exception handlers. 
    For instance, the merged 
    \pr{https://github.com/FederatedAI/FATE/pull/86}{86}
    of FATE points out a buggy \texttt{finally} statement. 
    \item {\bf Incompatible Dependency (IncDep)} refers to 
    incompatible dependencies of components and libraries, 
    e.g., 
    \pr{https://github.com/FederatedAI/FATE/pull/3787}{3787} finds that
    \texttt{mkdocstrings} specified in \texttt{doc/mkdocs/requirements.txt} is incompatible with others
    in FATE.
    \item {\bf Logic Error (LogicErr)} refers to incorrect, missing, or redundant code logic in source code or scripts.
    It has two primary sub-categories: \textit{Incorrect Logic} and \textit{Missing Logic}, where the former 
    refers to inaccurate code logic while the latter 
    refers to incomplete code logic. 
%
    \item {\bf Structure Error (StructErr)} indicates 
    an incorrect module structure of a framework, e.g., 
    \pr{https://github.com/OpenMined/PySyft/pull/3642}{3642} discovers that
    the lack of an \texttt{\_\_init\_\_.py} file in \texttt{syft/generic/abstract} leads to a top severity label. 
    \item {\bf Type Error (TypeErr)} refers to incorrect variable types and improper
    type conversions including both implicit (e.g., 
    \pr{https://github.com/OpenMined/PySyft/pull/5443/commits/2e2e40c0cd00996f828167fcff097274d02f5367}{5443}
    in PySyft) and explicit (e.g., 
    \pr{https://github.com/securefederatedai/openfl/pull/683}{683}
    in OpenFL) ones.
    \item {\bf Typo} is characterized by misspellings or errors in writing,
    e.g., 
    \pr{https://github.com/SymbioticLab/FedScale/pull/112}{112} discovers that
    {\tt download} and {\tt fedscale} are written as {\tt sownload}
    and {\tt FedScale} respectively in FedScale. 
 \end{enumerate} 

\begin{table}[t]
\centering \setlength{\tabcolsep}{4pt}
\caption{Distribution of bugs w.r.t. root causes.} 
\label{tab:root-cause-distribution}
\scalebox{.76}{ 
\begin{tabular}{lcc|lcc}
\toprule
{\bf Root Cause}                  & {\bf \#Bug}   & {\bf Percentage} & {\bf Root Cause}                    & {\bf \#Bug}   & {\bf Percentage} \\ 
\midrule
Logic Error (LogicErr)            & 418            & 37.4\%           & Assignment Problem (AsgProb)        & 318            & 28.4\%           \\
Condition Problem (CndProb)       & 124            & 11.1\%           & API Error (APIErr)                  & 97             & 8.7\%            \\
Typo                              & 61             & 5.5\%            & Documentation Error (DocErr)        & 52             & 4.6\%            \\
Type Error (TypeErr)              & 31             & 2.8\%            & Exception Handling Error (ExcHdErr) & 25             & 2.2\%            \\
Incompatible Dependency (IncDep)  & 19             & 1.7\%            & Computation Error (CompErr)         & 12             & 1.1\%            \\
Concurrency Error (CncyErr)       & 4              & 0.4\%            & Structure Error (StructErr)         & 2              & 0.2\%            \\
\bottomrule
\end{tabular}

}
\end{table}

Table~\ref{tab:root-cause-distribution} reports
the statistics of the root causes.
Logic Error and Assignment Problem are the two most frequent root causes, accounting for 37.4\% (418/1,119) and 28.4\% (318/1,119) of bugs, respectively. 
This is consistent with that of DL frameworks~\cite{ChenLSQJL23}.
The two most frequent sub-categories of Logic Error
are Incorrect Logic and Missing Logic, accounting for 71.3\% (298/418) and 23.9\% (100/418) of bugs, respectively. It indicates that it is more often to write
incorrect code than incomplete code.
The sub-categories of Assignment Problem, i.e., Assignment Error and Assignment Missing, account for 71.1\% (226/318) and 30.2\% (96/318) of bugs, respectively, where
four bugs 
reported in
\pr{https://github.com/FederatedAI/FATE/pull/76/commits/e1f2b46add678b6e375bb78d4bcb68df6a88c49d}{76},
\pr{https://github.com/FederatedAI/FATE/pull/4238/commits/d45c2c5b1951980f33171a463620fa48f0ee11a7}{4238},
\pr{https://github.com/primihub/primihub/pull/15/commits/790ef132c6d679a805743bd183a45609ee39fa7d}{15},
and \pr{https://github.com/primihub/primihub/pull/18}{18}
involve both sub-categories.
It reveals that a large number of bugs are caused by assignment errors.
  
 
Condition Problem is the third most frequent root cause, accounting for 11.1\% (124/1,119). 
Its two sub-categories, Condition Error and Condition Missing, account for 38.7\% (48/124) and 63.7\% (79/124) of bugs, respectively, where three bugs 
reported in 
\pr{https://github.com/FederatedAI/FATE/pull/1012/commits/e77942a6185c81cf35d62238f274927ce7599495}{1012},
\pr{https://github.com/FederatedAI/FATE/pull/1118/commits/256fcb2c139d6412ad964a9bf23cd069943aea5e}{1118}, and
\pr{https://github.com/FederatedAI/FATE/pull/3391/commits/3cc5d77ace52eee256d78e6c7402b5d9e8083be4}{3391}
involve both two sub-categories.
It is worth mentioning that this root cause has no counterpart in the studies of DL framework bugs~\cite{JiaZWHL21, SunZWDBC21, YangHXF22, DuSLA23, ChenLSQJL23},
highlighting the complexity of conditions that need to be considered and coped with in FL frameworks.

API Error is another frequent and notable root cause in FL frameworks, covering 8.7\% (97/1,119) of bugs. Its sub-categories, API Missing and API Misuse, contribute 43.3\% (42/97) and 56.7\% (55/97) of bugs, respectively.
While API Error is also a frequent and notable root cause in DL frameworks,
there is a slight difference between FL and DL frameworks.
Specifically, API Misuse contributes 80.0\% of bugs caused
by API Error, thus is significantly more frequent than API Missing in DL frameworks~\cite{ChenLSQJL23}.
This discrepancy indicates that the usage of APIs in FL frameworks is likely to be harder to keep in mind but less confusing than those in DL frameworks.

\begin{tcolorbox}[size=title]
    {\bf Finding 2:}
    {\it Logic Error} and {\it Assignment Problem} are the two most frequent root causes, caused 37.4\% (418/1,119) and 28.4\% (318/1,119) of bugs, respectively.
    {\it Condition Problem} is a notable root cause in FL frameworks, resulting in 11.1\% (124/1,119) of bugs. 
\end{tcolorbox} 

\smallskip
\noindent
{\bf Implication:} 
This finding indicates that a significant proportion (37.4\%+28.4\%+11.1\%+8.7\%=85.6\%) of bugs are caused by erroneous code logic, assignments, conditions and API usage. While an assignment or condition
or API usage error occurs in one line of code,  erroneous code logic and condition missing may involve multiple lines of code.
Consequently, the need arises for automated bug localization and fixing tools capable of effectively addressing issues spanning multiple lines of code.
On the other hand, most of these root causes come from developers' misunderstanding or misimplementation,
namely, inconsistency between the implementation and the expected behavior is their typical manifestation. Thus, a lightweight automated consistency checking tool could
be applied to prevent a large number of bugs caused by those root causes, following agile software development principles~\cite{beck2001manifesto, WilliamsC03}. This kind of tool should be able to utilize simple specifications to define the expected behavior and perform continuous checks throughout the development process.
Furthermore, equal attention should be paid to both API Missing and API Misuse
when detecting API errors in FL frameworks.

\smallskip
\noindent
{\bf Comparison with Root Causes in DL Frameworks.}
(1) Logic Error and Assignment Problem are also the two most frequent root causes in DL frameworks~\cite{ChenLSQJL23};
(2) Condition Problem and Documentation Error have no counterparts in the studies of DL frameworks~\cite{JiaZWHL21, SunZWDBC21, YangHXF22, DuSLA23, ChenLSQJL23};
and (3) API Missing and API Misuse contribute equally to API Error in FL framework bugs, in contrast, API Misuse is much more frequent than API Missing in DL frameworks~\cite{ChenLSQJL23}.

\begin{figure}[h]
  \centering
  \includegraphics[width=0.6\linewidth]{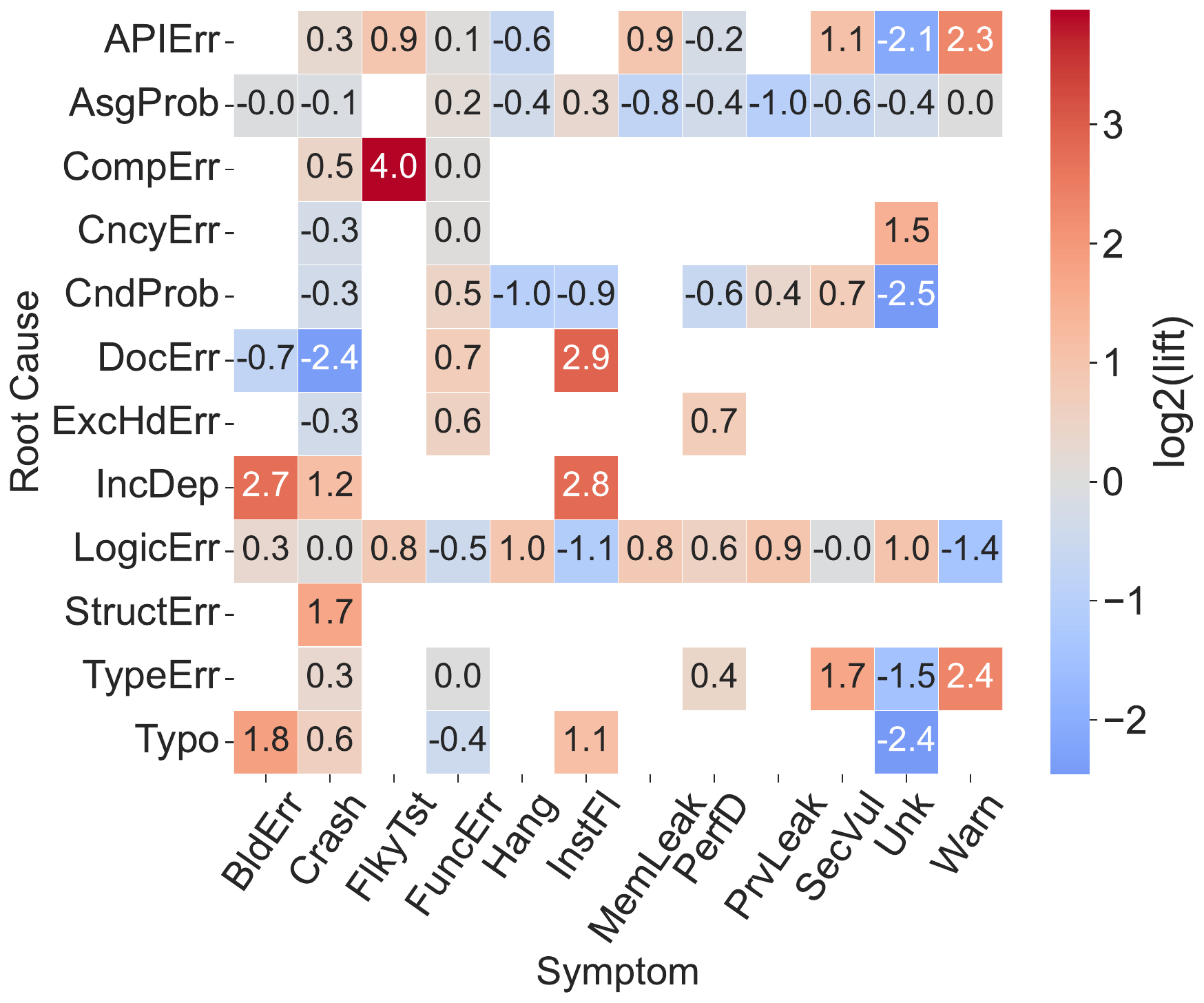}
  \caption{Correlation between symptoms and root causes in the terms of {\it lift} metric.}
  \label{fig:symptoms->root causes}
\end{figure}

\subsubsection{Correlation Between Root Causes and Symptoms}

Fig.~\ref{fig:symptoms->root causes} presents the correlation between root causes 
and symptoms in terms of the {\it lift} correlation metric.
Since 
the \textit{lift} values of some pairs are significantly higher than others,
all the \textit{lift} values are logarithmized for better exhibition, i.e., $\log_2(\textit{lift}(A, B))$. 
Furthermore, the grids of pairs $(A,B)$ with the strongest negative correlation (i.e., $\textit{lift}(A,B)=0$) in the heat maps are left empty.

Interestingly, the two most frequent symptoms \textit{Functional Error} and \textit{Crash} are not strongly, positively correlated with the
two most frequent root causes {\it Logic Error} and {\it Assignment Problem}.
Instead, \textit{Functional Error} is more positively correlated with less frequent root causes: {\it Documentation Error, Exception Handling Error}, and {\it Condition Problem}.
We found that errors in documentations can mislead users to write incorrect programs which lead to unexpected results. For example, 
\issue{https://github.com/FederatedAI/FATE/issues/3522}{3522} reports that
the develop documentation does not mention that the \texttt{fate-flow} service needs to be restarted
after modifying parameters of the model, causing a functional error. 
{\it Exception Handling Error} and {\it Condition Problem} misdirect  control flow
and lead to unexpected results.
For instance, 
\issue{https://github.com/FederatedAI/FATE/issues/85}{85} reports that
an incorrect \texttt{finally} branch in FATE leads to a cleaning up operation at the wrong time, 
and 
\issue{https://github.com/OpenMined/PySyft/issues/3214}{3214} discovers that
an erroneous condition branch which instantiates a \texttt{Placeholder} with a wrapper tensor in PySyft leads to incorrect parameter serialization results. 

\begin{figure}[t]
    \centering
\begin{lstlisting}[
        language=Python, 
        label=lst:CompErr2Crash, 
        caption={When \texttt{self.percentiles}=$1$, \texttt{thres\_idx} is \texttt{len(sorted\_value)}, which is out of the bound}.
    ]
# federatedml/feature/feature_selection.py, class UnionPercentileFilter(FilterMethod)
def get_value_threshold(self):
    ...
    thres_idx = int(math.floor(self.percentiles * len(sorted_value)))
    self.value_threshold = sorted_value[thres_idx]
\end{lstlisting}\vspace{-1mm}
\end{figure}

Similarly, \textit{Crash} is more positively correlated with less frequent root causes: {\it API Error}, {\it Computation Error},
{\it Incompatible Dependency}, {\it Structure Error}, {\it Type Error},
and {\it Typo}. We found when such an error occurs, 
some exception is raised either explicitly or implicitly,
but is not handled properly, thus resulting in a crash.  
For instance, a computation error 
reported in \issue{https://github.com/FederatedAI/FATE/issues/286}{286}
results in an incorrect index {\tt thres\_idx}
and then a crash because the index is out of the bound and the exception is not handled properly as shown in Code Snippet~\ref{lst:CompErr2Crash};
the missing of importing the \texttt{FloatTensor} class
discovered in \pr{https://github.com/OpenMined/PySyft/pull/1315}{1315}
leads to an unhandled \texttt{NameError} exception in PySyft; 
and the missing of the \texttt{bind\_protobuf} decorator for protobuf serialization results in an unexpected \texttt{TypeError} exception 
reported by \issue{https://github.com/OpenMined/PySyft/issues/5208}{5208}
in PySyft. 

\textit{Privacy Leak}
is positively correlated with \textit{Logic Error}
and \textit{Condition Problem}.
Unsurprisingly, \textit{Logic Error} can cause privacy leakage.
For instance, 
\issue{https://github.com/FederatedAI/FATE/issues/869}{869} finds that
the missing of encryption in FATE leads to the unintended leakage of the private intermediate loss values. 
Remarkably, the correlation with \textit{Condition Problem} indicates that errors in conditions or branching statements can also undermine data privacy,
e.g., 
\pr{https://github.com/FederatedAI/FATE/pull/3391}{3391} in FATE discovers that
the anonymous in the host party is revealed due to the missing of a condition checking. 

From the perspective of root causes, 
the most frequent root cause,
\textit{Logic Error}, is positively correlated with various symptoms:
\textit{Build Error}, \textit{Flaky Test}, \textit{Hang}, \textit{Memory Leak}, \textit{Performance Degradation}, and \textit{Privacy Leak}.
For instance, 
\issue{https://github.com/FederatedAI/FATE/issues/3910}{3910} in FATE reports a Build Error caused by the missing of reading an environment variable;
\issue{https://github.com/OpenMined/PySyft/issues/3359}{3359} discovers that 
an incorrect assumption in PySyft leads to improper serialization strategies
which results in inconsistent and unreliable test results; 
\issue{https://github.com/NVIDIA/NVFlare/issues/193}{193} finds that
incorrect ordering in NVFlare results in messages being transmitted at the wrong time which prevents clients from shutting down; 
\issue{https://github.com/OpenMined/PySyft/issues/3316}{3316} in PySyft discovers memory leakage caused by improper handling of intermediate tensors;
\issue{https://github.com/tensorflow/federated/issues/896}{896} reports that
erroneous dataset decompression increases the execution time significantly in TFF; 
and improper selection of trainers and hyperparameters in PySyft leads to abnormally high loss values, which was found by 
\issue{https://github.com/OpenMined/PySyft/issues/2642}{2642}. 
Furthermore, \textit{Logic Error} is the dominant root cause of \textit{Hang}.

In contrast, the second frequent root cause, 
\textit{Assignment Problem}, is correlated with many symptoms, but
is only weakly positively correlated with \textit{Functional Error}
and \textit{Installation Failure}. This indicates
that \textit{Assignment Problem} can cause many different symptoms, but is not a major
root cause of specific symptoms. 

Last but not least, (1) \textit{API Error},  
\textit{Condition Problem}, {\it Incompatible Dependency}, and \textit{Type Error}
can cause at least three distinct symptoms; (2) 
{\it Documentation Error}, {\it Incompatible Dependency} and {\it Typo} are major root causes
of {\it Build Error} and {\it Installation Failure};
(3) {\it Flaky Test} is mainly caused by {\it API Error}, {\it Computation Error} and {\it Logic Error};
(4) {\it Warning} is often caused by {\it API Error} and \textit{Type Error};
(5) bugs with  \textit{Unknown} symptom are typically caused
by \textit{Logic Error} and \textit{Concurrency Error}.
\begin{tcolorbox}[size=title]
    {\bf Finding 3:}
    The two most frequent symptoms, \textit{Functional Error} and \textit{Crash}, 
    are mainly caused by less frequent root causes:
    {\it Documentation Error, Exception Handling Error}, and {\it Condition Problem}.
    The two most frequent root causes, \textit{Logic Error} and  \textit{Assignment Problem},
    can lead to diverse symptoms such as \textit{Build Error}, \textit{Flaky Test}, \textit{Hang}, \textit{Memory Leak}, \textit{Performance Degradation}, \textit{Privacy Leak}, \textit{Functional Error},
and \textit{Installation Failure}.
  The FL-specific symptom \textit{Privacy Leak} is often caused by \textit{Logic Error} and \textit{Condition Problem}. 
\end{tcolorbox}

\smallskip
\noindent
{\bf Implication:} 
This finding indicates that (1) detection and localization of functional error and crash should consider various root causes and test cases should be diverse;
(2) more attentions should be paid to exceptions to address crash bugs;
(3) logic error and condition problem should be the focus of detection and localization of privacy leakage;
and (4) 
path-sensitive analysis tools have the potential to efficiently and effectively detect functional error, crash, security vulnerability, and privacy leakage.

\smallskip
\noindent
{\bf Comparison with Correlation between Root Causes and Symptoms in DL Frameworks.}
(1) while functional errors are positively correlated with logic errors in DL frameworks~\cite{JiaZWHL21}, they are negatively correlated in FL frameworks;
and (2) although incompatible dependency never causes warnings in FL frameworks, it does in DL frameworks~\cite{JiaZWHL21}.

\subsection{RQ3: Fix Patterns and their Correlation to Root Causes}
\subsubsection{Fix Patterns}

To comprehensively understand fix patterns, we identify six categories
of fix patterns, each of which is a location of the modified objects 
for fixing bugs. As shown in Table~\ref{tab:fix-pattern-distribution}, some category is further divided into one or more sub-categories according to the root causes. 

\begin{enumerate}[leftmargin=*]
    \item {\bf Comment (Cmt)} refers to the fix pattern that updates comments in the source code to improve readability and eliminates misunderstandings.
    For instance, in \pr{https://github.com/alibaba/FederatedScope/pull/223}{223}, a comment that describes how to disable early stop in the \texttt{extend\_training\_cfg} method in FederatedScope is updated. 
%
    \item {\bf Configuration (Cfg)} refers the pattern that fixes  key-value pairs 
    (e.g., the incorrect version of \texttt{flwr} in \texttt{pyproject.toml} in \pr{https://github.com/adap/flower/pull/1344}{1344})~\cite{Cfg_FKVPrs})
    and typos (e.g., the typo 
    \pr{https://github.com/FederatedAI/FATE/pull/2129/commits/48d27e589ac3dbf462e64e630b745ef22a5106d0}{2129}
    in a \texttt{JSON} file that configures job and module settings for all participants in a FL task) 
    in configuration files. 
    \item {\bf Documentation (Doc)} refers to the pattern that rectifies incorrect or update obsolete information or instructions in documentations that mislead users.
    (e.g., 
    \pr{https://github.com/OpenMined/PySyft/pull/203}{203} in PySyft).
%
    \item {\bf Implementation (Impl)} 
    refers to the pattern that fixes bugs by modifying source code. 
    It is divided into eight sub-categories, as shown in Table~\ref{tab:fix-pattern-distribution}.
%
    \item {\bf Script (Scr)} consists of five sub-categories, as shown in Table~\ref{tab:fix-pattern-distribution}, all of which fix 
    bugs in scripts. 
%
    \item {\bf Structure(Struct)} is the pattern that 
    reorganizes module structures by adding new files to fix 
    the structure errors 
    (e.g., 
    \pr{https://github.com/OpenMined/PySyft/pull/3642}{3642}
    in PySyft and 
    \pr{https://github.com/FederatedAI/FATE/pull/1759/commits/e3fdeb6dc840ff7be8216429408ca7ece3557c50}{1759}
    in FATE). 
\end{enumerate}

\begin{table}[t]
\centering\setlength{\tabcolsep}{4pt}
\caption{Distribution of bugs w.r.t. fix patterns.}
\label{tab:fix-pattern-distribution}
\scalebox{.75}{
\begin{tabular}{lcc|lcc}
\toprule
\textbf{Fix Pattern}          & \textbf{\#Bug}  & \textbf{Percentage} & \textbf{Fix Pattern}                  & \textbf{\#Bug}  & \textbf{Percentage} \\ 
\midrule
\textbf{Comment}              & \textbf{10}  & \textbf{0.9\%}      & \textbf{Implementation}                       & \textbf{911} & \textbf{81.4\%}     \\
Update Comment (UdCmt)        & 10           & 0.9\%               & Add Type Conversion or Fix Type (ATpConv/FTp) & 25           & 2.2\%               \\ \cline{1-3}
\textbf{Configuration}        & \textbf{123} & \textbf{11.0\%}     & Fix APIs (FAPI)                               & 127          & 11.3\%              \\
Fix Key-Value Pairs (FKVPrs)  & 118          & 10.5\%              & Fix Assignment (FAsg)                         & 209          & 18.7\%              \\
Fix Typo (FTypo)              & 6            & 0.5\%               & Fix Calculation Expression (FCalcExpr)        & 13           & 1.2\%               \\ \cline{1-3}
\textbf{Script}               & \textbf{42}  & \textbf{3.8\%}      & Fix Code Logic (FCodLog)                      & 377          & 33.7\%              \\
Fix APIs (FAPI)               & 1            & 0.1\%               & Fix Condition Blocks (FCndBlk)                & 125          & 11.2\%              \\
Fix Assignment (FAsg)         & 18           & 1.6\%               & Fix Exception Handling Blocks (FExcBlk)       & 22           & 2.0\%               \\
Fix Code logic (FCodLog)      & 20           & 1.8\%               & Fix Typo (FTypo)                              & 32           & 2.9\%               \\ \cline{4-6}
Fix Condition Branch (FCndBr) & 1            & 0.1\%               & \textbf{Documentation}                        & \textbf{63}  & \textbf{5.6\%}      \\  
Fix Typo (FTypo)              & 4            & 0.4\%               & Update Information (UdInfo)                   & 63           & 5.6\%               \\ \cline{1-3}
\textbf{Structure}            & \textbf{2}   & \textbf{0.2\%}      &                                               &              &                     \\
Add New File (ANF)            & 2            & 0.2\%               &                                               &              &                     \\ \bottomrule
\end{tabular}%

}
\end{table}

Table~\ref{tab:fix-pattern-distribution} shows the statistics of the fix patterns.
The largest number of bugs are fixed by modifying source code, covering 81.4\% (911/1,119) of bugs, among which 41.4\% (377/911), 22.9\% (209/911), 13.9\% (127/911), and 13.7\% (125/911) are fixed by modifying code logic, assignments 
(e.g., \pr{https://github.com/scaleoutsystems/fedn/pull/277}{277} in FEDn 
and \pr{https://github.com/alibaba/FederatedScope/pull/223}{223} in FederatedScope),
APIs 
(e.g., \pr{https://github.com/OpenMined/PySyft/pull/3356}{3356},
\pr{https://github.com/OpenMined/PySyft/pull/1315}{1315},
and \pr{https://github.com/OpenMined/PySyft/pull/3525}{3525} 
in PySyft),
and condition blocks 
(e.g., \pr{https://github.com/OpenMined/PySyft/pull/4801}{4801}
and \pr{https://github.com/OpenMined/PySyft/pull/3442}{3442}
in PySyft),
respectively.
We found that 19 bugs 
are fixed by multiple fix patterns that modify source code
(e.g., \pr{https://github.com/OpenMined/PySyft/pull/4750/commits/cd7a38e9d7fbc0590b601a848794ef6cf75762d2\#diff-107dbd8c9752804a036e625706eb5f2d4e4791e338be8fd523c5f75649562cff}{4750}
in PySyft
and \pr{https://github.com/alibaba/FederatedScope/pull/210}{210}
in FederatedScope).
This distribution is in general consistent with that of the root causes, namely, {\it Logic Error}, {\it Assignment Problem}, 
{\it Condition Problem}, and {\it API Error} are the four most frequent root causes.
It indicates that fix patterns in source code play a dominant role in bug fixing.
 
The second largest number of bugs are fixed by modifying configurations, accounting for 11.0\% (123/1,119) of bugs.
Remarkably, 95.9\% (118/123) of them are fixed by 
modifying key-value pairs of configurations. 
It is because FL frameworks usually have a number of configurations for build, installation and setting of FL tasks.
For example, FATE uses \texttt{JSON} to describe FL tasks,
model parameters and participants~\cite{FATEConfSys}, where
data-transform, feature-engineering, and classification/regression module are combined as a directed acyclic graph. 
This indicates that configurations in FL frameworks are bug-prone and such bugs
are fixed by revising the key-value pairs.

A notable number of bugs are fixed by modifying documentations, accounting for 5.6\% (63/1,119) bugs. It may be attributed to the early stage of most FL
frameworks so that documentations are unclear and misleading.
We also note that most of bugs in scripts are fixed by modifying assignments 
(e.g., \pr{https://github.com/OpenMined/PySyft/pull/2278/commits/01fb99d8b8d5307bf0a9eea7425396e761d2cae5\#diff-76ed074a9305c04054cdebb9e9aad2d818052b07091de1f20cad0bbac34ffb52}{2278}
in PySyft) 
and code logic 
(e.g., \pr{https://github.com/FederatedAI/FATE/pull/2129/commits/48d27e589ac3dbf462e64e630b745ef22a5106d0}{2129}
in FATE).
Only one bug is fixed by both fix patterns in 
\pr{https://github.com/FederatedAI/FATE/pull/1300/commits/52fbf0cf3ad1b6f62ff6abc15ffcd4d4df0ef1af}{1300}
in FATE.

\begin{tcolorbox}[size=title]
    {\bf Finding 4:} 
    The majority of bugs are fixed in source code implementation, accounting for 81.4\% (911/1,119) of bugs, most of which fix code logic, assignment, APIs, and conditional branches.
    The two most frequent fix patterns in scripts are also related to code logic and assignment. 
\end{tcolorbox} 

\smallskip
\noindent
{\bf Implication:} 
This finding indicates that 
(1) the ability to fix flaws in source code is undoubtedly necessary for automated bug fixing tools; 
and (2) 
such ability should be extended to address errors in scripts
and  erroneous key-value pairs in configuration files by making necessary adaptions.
The former has been widely studied in the literature in the other domains\cite{WenMing2018,Koyuncu2018FixMinerMR,Xia2022LessTM}. However, the latter is less considered
and thus is interesting future work.

\smallskip
\noindent
{\bf Comparison with Bug Fix Patterns in DL Frameworks}.
(1) The counterparts of modifying code logic and assignments are also major bug fix patterns in DL frameworks, addressing~24.4\% of bugs~\cite{SunZWDBC21}, and 34.5\% of bugs~\cite{JiaZWHL21}, respectively;
(2)~11.4\% of bugs in FL frameworks are fixed by modifying APIs, 
while~28.6\% of bugs are fixed by its counterparts in DL frameworks~\cite{JiaZWHL21};
(3)~11.3\% of bugs in FL frameworks are addressed by modifying condition expressions and conditional branches, 
while its counterparts fix~21.9\% and~29.1\% of bugs in DL frameworks in~\cite{JiaZWHL21} and ~\cite{SunZWDBC21}, respectively;
and (4)~2.2\% of bugs are fixed by modifying types of variables in FL frameworks, significantly less than that (11.8\%) in DL frameworks.

\subsubsection{Correlation Between Fix Patterns and Root Causes}

\begin{figure}
  \centering
  \includegraphics[width=0.6\columnwidth]{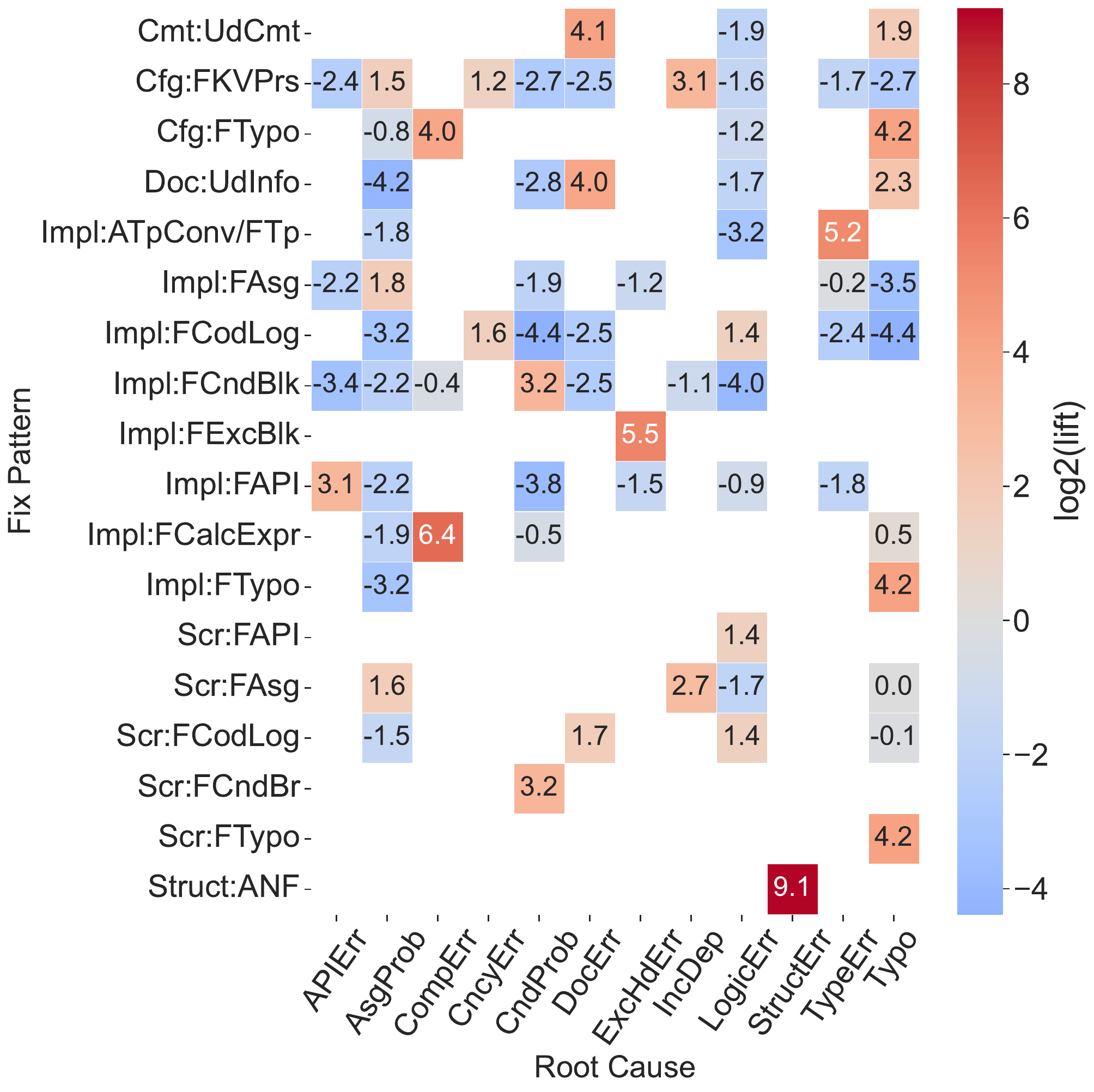}
  \caption{Transformed \textit{lift} correlation values between root causes and fix patterns.}
  \label{fig:root causes->fix patterns}
\end{figure}

Fig.~\ref{fig:root causes->fix patterns} depicts the correlation between root causes and fix patterns.
We remark that that many fix patterns are named specifically to address bugs caused by their corresponding root causes.
Consequently, the \textit{lift} values measuring the correlation between root causes and fix patterns tend to be significant. 

\begin{figure}
    \centering
\begin{lstlisting}[
        language=diff, 
        label=lst:ScrFAPI2LogicErr, 
        caption=The new  class \texttt{DevelopGRPC} is added to fix a missing necessary gRPC build step.
    ]
  # setup.py
+ class DevelopGRPC(develop):
+     """Command for develop installation."""
+     def __init__(self, dist):
+         """Create a sub-command to execute."""
+         self.subcommand = BuildPackageProtos(dist)
+         super().__init__(dist)
+     def run(self):
+         """Build GRPC modules before the default installation."""
+         self.subcommand.run()
+         super().run()
\end{lstlisting}\vspace{-3mm}
\end{figure}

\textit{Logic Error}, as the most frequent root cause, is positively correlated with not only \textit{Fix Code Logic} in source code and scripts, but also \textit{Fix APIs} in scripts because of one bug reported in \issue{https://github.com/securefederatedai/openfl/issues/394}{394} in OpenFL.
This bug results from the missing logic for a necessary gRPC build step
and is fixed by adding a new class \texttt{DevelopGRPC} in \texttt{setup.py} script as shown in Code Snippet~\ref{lst:ScrFAPI2LogicErr} 
from \pr{https://github.com/securefederatedai/openfl/pull/395}{395}.

In contrast, \textit{Assignment Problem}, as the second most frequent root cause, is correlated with multiple fix patterns,
but is only positively correlated with fixing assignments in source code, scripts, and configuration files. 
This indicates that bugs caused by \textit{Assignment Problem} can be addressed by various fix patterns, but none of them is a major solution, except for modifying assignments directly in source code, scripts, and configuration files.
Similarly, \textit{Condition Problem} 
is correlated with various fix patterns, but is only positively correlated with modifications on condition expressions and conditional branches in source code and scripts.



From the perspective of fix patterns, the four most frequent fix patterns, which all modify source code, are correlated with 
at least six distinct root causes. 
However, each of the four fix patterns is only positively correlated with one root cause, which corresponds to the definition of the fix pattern.

Lastly, 
(1) the following fix patterns are positively correlated with every root cause they have been utilized to fix and serve as major solutions: fixing typos in configuration files and scripts, modifying exception handling in source code, fixing APIs in scripts, and adding new files to the module structure;
(2) updating code comments and documentation are the major solutions to \textit{Documentation Error} and \textit{Typo};
(3) fixing key-value pairs in configuration files and modifying code logic in source code are the  major solutions to \textit{Concurrency Error};
(4) fixing typos and incorrect calculation expressions in source code are the the major solutions to \textit{Computation Error};
and (5) besides modifying assignments in scripts, fixing key-value pairs in configuration files is another major solution to \textit{Incompatible Dependency}.

\begin{tcolorbox}[size=title]
    {\bf Finding 5:}
    The three most frequent root causes, \textit{Logic Error}, \textit{Assignment Problem}, and \textit{Condition Problem}, are majorly fixed by the respective fix patterns that are named to address them.
    The four most frequent fix patterns are positively correlated with only the root causes they are named to address.
    In scripts, the two most frequent fix patterns, \textit{Fix Code Logic} and \textit{Fix Assignment}, have strong positive correlations with not only their respective root causes, but also \textit{Documentation Error} and \textit{Incompatible Dependency}, respectively.
\end{tcolorbox} 

\smallskip
\noindent
{\bf Implication:} 
This finding indicates that 
(1) fix patterns, which directly targets the respective root causes, are unsurprisingly the most effective solutions to bugs;
and (2) attention is needed to fix bugs in scripts when errors occur in documentation and configurations for dependencies.

\subsection{RQ4: Cross Analysis of Distributions across Functionalities}

\subsubsection{Functionalities}

We establish a taxonomy for main functionalities according to the similarities and differences of architecture designs of FL frameworks. 
Each functionality implements a specific function and covers related tasks that are special or necessary in FL frameworks.

\begin{table}[t]
\centering \setlength{\tabcolsep}{4pt}
\caption{Distribution of bugs w.r.t. functionalities.}
\label{tab:functionalities-distribution}
\resizebox{0.99\columnwidth}{!}{%
\begin{tabular}{@{}clcc|clcc@{}}
\toprule
\multicolumn{2}{c}{\textbf{Functionality}}                                  & \textbf{\#Bug} & \textbf{Percentage} & \multicolumn{2}{c}{\textbf{Functionality}}                                               & \textbf{\#Bug} & \textbf{Percentage} \\ 
\midrule
\multicolumn{1}{c|}{\multirow{12}{*}{Core}} & Algorithm (Algo)              & 117         & 10.5\%              & \multicolumn{1}{c|}{\multirow{4}{*}{Core}}       & Privacy Preservation (PrvPrsvn)       & 34          & 3.0\%               \\
\multicolumn{1}{c|}{}                       & Anomaly Detection (AnmlyDet)  & 2           & 0.2\%               & \multicolumn{1}{c|}{}                            & Resource Management (RsrcMgmt)        & 7           & 0.6\%               \\
\multicolumn{1}{c|}{}                       & Data Collection (DataColl)    & 41          & 3.7\%               & \multicolumn{1}{c|}{}                            & Security Protection (SecProt)         & 9           & 0.8\%               \\
\multicolumn{1}{c|}{}                       & Data Preprocessing (DataPrep) & 67          & 6.0\%               & \multicolumn{1}{c|}{}                            & Training Management (TrngMgmt)        & 17          & 1.5\%               \\
\cline{5-8} 
\multicolumn{1}{c|}{}                       & Data Storage (DataSTG)        & 37          & 3.3\%               & \multicolumn{1}{c|}{\multirow{7}{*}{Auxiliary}}  & Build \& Installation (Bld\&Inst)     & 51          & 4.6\%               \\
\multicolumn{1}{c|}{}                       & Data Transmission (DataXmsn)  & 21          & 1.9\%               & \multicolumn{1}{c|}{}                            & Documentation (Doc)                   & 58          & 5.2\%               \\
\multicolumn{1}{c|}{}                       & Evaluation (Eval)             & 26          & 2.3\%               & \multicolumn{1}{c|}{}                            & Example                               & 145         & 13.0\%              \\
\multicolumn{1}{c|}{}                       & Feature Engineering (FeatEng) & 29          & 2.6\%               & \multicolumn{1}{c|}{}                            & Repository Management (RepoMgmt)      & 13          & 1.2\%               \\
\multicolumn{1}{c|}{}                       & Inference (Infrnc)            & 5           & 0.4\%               & \multicolumn{1}{c|}{}                            & System Utility (SysUtil)              & 335         & 29.9\%              \\
\multicolumn{1}{c|}{}                       & Messaging (Msging)            & 45          & 4.0\%               & \multicolumn{1}{c|}{}                            & Testing                               & 97          & 8.7\%               \\
\multicolumn{1}{c|}{}                       & Model Aggregation (ModAggr)   & 19          & 1.7\%               & \multicolumn{1}{c|}{}                            & Workflow Design (WfDsgn)              & 165         & 14.7\%              \\
\multicolumn{1}{c|}{}                       & Model Training (ModTrng)      & 39          & 3.5\%               & \multicolumn{1}{c|}{}                                                 &                                       &             &                     \\ 
\bottomrule
\end{tabular}

}
\end{table}

Our taxonomy of main functionalities encompasses a total of 23 categories, which are further grouped into two parts: the {\bf core} functionalities, consisting of functionalities adapted from existing work~\cite{LoLWPZ21}, and the {\bf auxiliary} functionalities proposed by us.

Table~\ref{tab:functionalities-distribution} presents the statistics of the 23 main functionalities.
The core part consists of functionalities that are essential or specific to FL, involved in  40.8\% (457/1,119) of bugs.
On the other hand, the auxiliary part comprises supportive functionalities that are used to enhance user experience and accelerate the development process, involved in 65.8\% (736/1,119) of bugs.
We remark that 74 bugs involve both the core and auxiliary functionalities 
(e.g., \issue{https://github.com/TL-System/plato/issues/102}{102} 
and \issue{https://github.com/TL-System/plato/issues/174}{174}
in Plato).

 Among the core functionalities, 
\textit{Algorithm} is the most bug-prone one, involved in~10.5\%~(117/1,119) of bugs. 
It mainly covers  
the implementation of algorithms in various parties  
in FL frameworks, such as the roles of the clients and the central server (e.g., \issue{https://github.com/FederatedAI/FATE/issues/869}{869} in FATE). {\it Data Preprocessing}, {\it Messaging} 
and {\it Data Collection} are three notable
types of core functionalities, involved in~6.0\%~(67/1,119), 4.0\%~(45/1,119) and 3.7\% (41/1,119) of bugs. 

Among the auxiliary functionalities, 
\textit{System Utility} that relates to supportive classes and functions in FL frameworks, is the most bug-prone one, involved in~29.9\%~(335/1,119) of bugs. 
It includes command parsers and other various utilities that contribute to the overall usability of FL frameworks, but are loosely related to FL
(e.g. \issue{https://github.com/FedML-AI/FedML/issues/206}{206} in FedML). 
\textit{Workflow Design} that comprises classes and functions specifically implemented to enhance user-friendliness of FL workflow
(e.g., \pr{https://github.com/OpenMined/PySyft/pull/3383}{3383} in PySyft), is the second most bug-prone functionality, involved in 14.8\% (165/1,119) of bugs. 
\textit{Example} provides executable demo code, allowing users to gain 
familiarity with the usage of FL frameworks
(e.g., \issue{https://github.com/OpenMined/PySyft/issues/2642}{2642} in PySyft). 
As the third most bug-prone functionality, it is involved in 13.0\% (145/1,119) of bugs.
\begin{tcolorbox}[size=title]
    {\bf Finding 6:}
    Auxiliary functionalities are the most bug-prone functionalities,
    involved in 65.8\% (736/1,119) of bugs.
    The four most bug-prone functionalities are \textit{System Utility}, \textit{Workflow Design}, \textit{Example}, and \textit{Algorithm}, involved in 29.9\% (335/1,119), 14.7\% (165/1,119), 13.0\% (145/1,119), and 10.5\% (117/1,119) of bugs, respectively.
\end{tcolorbox} 

\smallskip
\noindent
{\bf Implication:} 
This finding suggests that more attention should be paid 
on developing auxiliary functionalities. 
Effective bug detection and fixing techniques 
for general systems software are likely to exhibit efficacy in 
\textit{System Utility} with domain-specific adaptions.
Similarly, tools that are used 
for DL core functionalities are also likely to be effective in FL core functionalities with specific adaptions.
Lastly, we recommend that developers conduct thorough code review and testing on the code examples provided in FL frameworks.



\begin{figure}[t]
  \centering
  \includegraphics[width=1\linewidth]{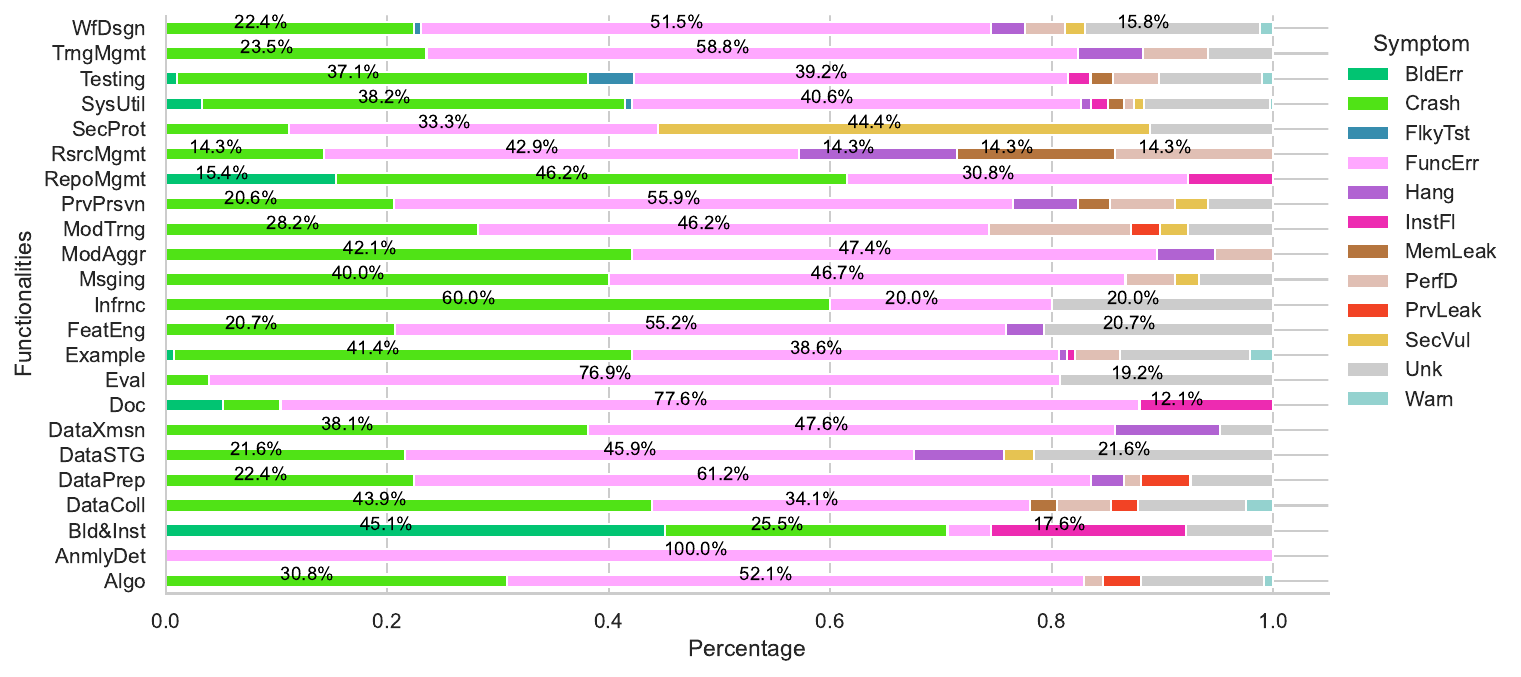}
  \caption{The distribution of symptoms across functionalities.}
  \label{fig:symptom-on-functionalities}
\end{figure}

To delve deeper into the classification results and their implications, we further investigate the distributions of symptoms, root causes, and fix patterns across the 23 main functionalities, which reveals the 
significance of the functionalities for bug detection, localization and repair.

\subsubsection{Distributions of Symptoms across Functionalities}

Fig.~\ref{fig:symptom-on-functionalities} depicts the distributions of symptoms across functionalities.
\textit{Functional Error} and \textit{Crash} are the two most frequent symptoms in almost all the functionalities, except for \textit{Build~\& Installation}, \textit{Security Protection}, and \textit{Evaluation}.
In \textit{Build~\& Installation}, \textit{Build Error} 
(e.g., \pr{https://github.com/OpenMined/PySyft/pull/2868}{2868}
in PySyft) 
and \textit{Crash} 
(e.g., \issue{https://github.com/NVIDIA/NVFlare/issues/736}{736}
in NVFlare) 
are the two most frequent symptoms, 
accounting for 45.1\% (23/51) and 25.5\% (13/51), respectively. 
In \textit{Security Protection}, the two most frequent symptoms are \textit{Security Vulnerability} and \textit{Functional Error} 
(e.g., \issue{https://github.com/FederatedAI/FATE/issues/689}{689}
in FATE).
In \textit{Evaluation}, \textit{Unknown} is more frequent 
than \textit{Crash}, and becomes the second most frequent symptom, accounting for 19.2\% (5/26) of bugs.

\textit{Privacy Leak},  the symptom that is unique to FL frameworks, occurs in four functionalities, namely, \textit{Data Preprocessing} 
(e.g., \pr{https://github.com/FederatedAI/FATE/pull/4106}{4106}
in FATE), 
\textit{Algorithm} 
(e.g., \issue{https://github.com/TL-System/plato/issues/97}{97}
in Plato), 
\textit{Model Training} 
(e.g., \issue{https://github.com/FederatedAI/FATE/issues/869}{869}
in FATE), 
and \textit{Data Collection} 
(e.g., \issue{https://github.com/TL-System/plato/issues/97}{97}
in Plato).
It indicates that these functionalities
are particularly susceptible to data privacy leakage. 

Bugs with an unknown symptom occur most frequently in the following five functionalities: {\it Data Storage, Feature Engineering, Inference, Evaluation}, and \textit{Workflow Design},
accounting for 21.6\% (8/37), 20.7\% (6/29), 20.0\% (1/5), 19.2\% (5/26), and 15.8\% (26/165) of bugs, respectively.



\begin{tcolorbox}[size=title]
    {\bf Finding 7:}
\textit{Functional Error} and \textit{Crash} are still the two most frequent symptoms across functionalities except for \textit{Build~\& Installation}, \textit{Security Protection}, and \textit{Evaluation}.
    Functionalities related to data collection, preprocessing, and training FL models, are particularly susceptible to \textit{Privacy Leak}.
    Five functionalities, i.e., {\it Data Storage, Feature Engineering, Inference, Evaluation}, and \textit{Workflow Design}, are prone to bugs with unknown symptoms.
\end{tcolorbox}

\smallskip
\noindent
{\bf Implication:} 
This finding indicates that
(1) privacy leakage analysis should focus more on the functionalities that involve data collection, data preprocessing, and FL model training;
and (2) strict bug report formats are particularly needed for
the functionalities that involve {\it Data Storage, Feature Engineering, Inference, Evaluation}, and \textit{Workflow Design}.

\begin{figure}[t]
  \centering
  \includegraphics[width=1\linewidth]{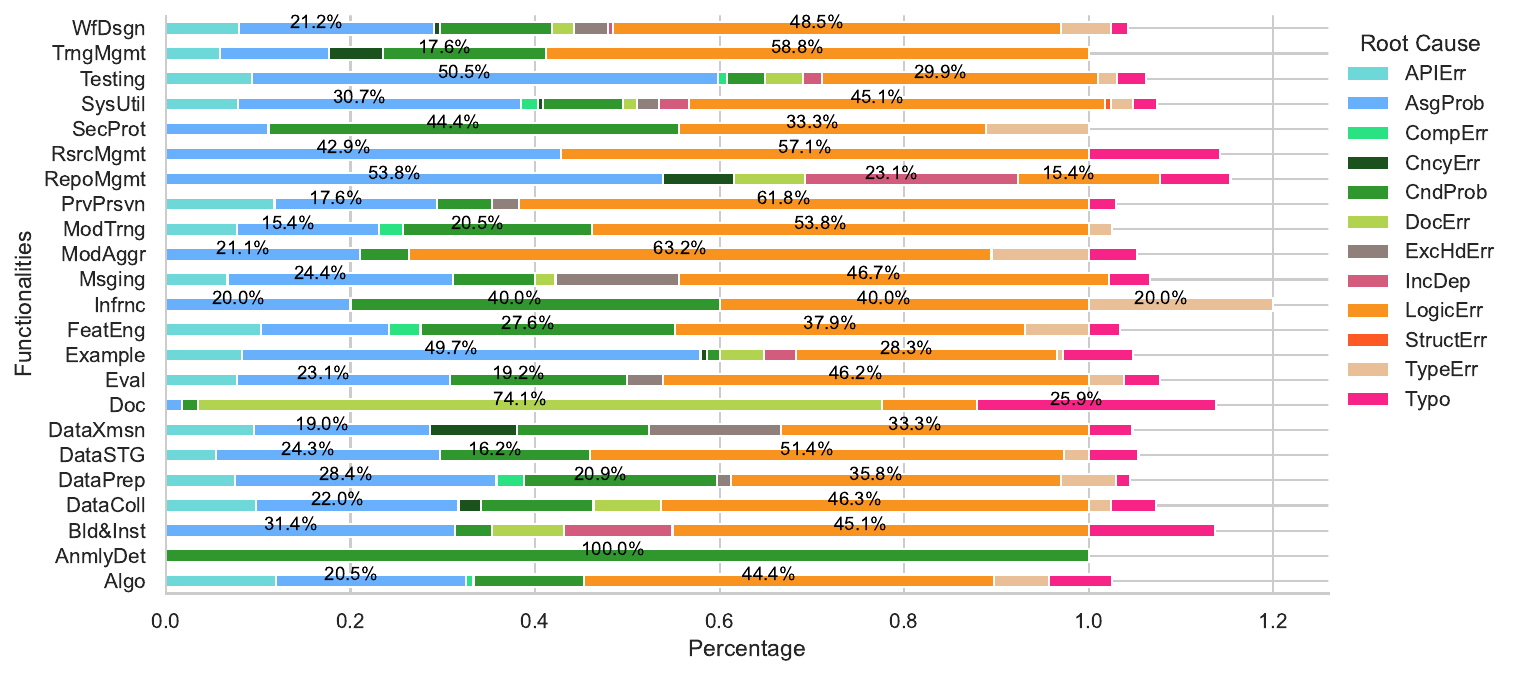}
  \caption{The distributions of root causes across functionalities.}
  \label{fig:root-cause-on-functionalities}
\end{figure}

\subsubsection{Distributions of Root Causes across Functionalities}

Fig.~\ref{fig:root-cause-on-functionalities} show the distributions of root causes across functionalities.
\textit{Logic Error} and \textit{Assignment Problem} are the two most frequent root causes in most functionalities except for {\it Documentation} and \textit{Repository Management}, where \textit{Logic Error} takes the place of the third most frequent root cause.
In \textit{Documentation}, it is unsurprising to find that the most two frequent root causes are 
\textit{Documentation Error} (e.g., \issue{https://github.com/OpenMined/PySyft/issues/125}{125} in PySyft) 
and \textit{Typo} (e.g., \pr{https://github.com/FederatedAI/FATE/pull/1030}{1030} in FATE), 
accounting for 74.1\% (43/58) and 25.9\% (15/58) of bugs, respectively.
In \textit{Repository Management}, the most two frequent root causes are 
\textit{Assignment Problem} (e.g., \issue{https://github.com/OpenMined/PySyft/issues/4927}{4927} in PySyft) 
and \textit{Incompatible Dependency} (e.g., \pr{https://github.com/securefederatedai/openfl/pull/517}{517} in OpenFL), accounting for 53.8\% (7/13) and 23.1\% (3/13) of bugs, respectively.
In addition, \textit{Condition Problem} dominates in \textit{Anomaly Detection}, \textit{Security Protection}, and \textit{Inference}, covering 100\% (2/2), 44.4\% (4/9), and 40.0\% (2/5) of bugs, respectively.


\begin{tcolorbox}[size=title]
    {\bf Finding 8:}
    \textit{Logic Error} and \textit{Assignment Problem} are still the two most frequent root causes across functionalities except for {\it Documentation} and \textit{Repository Management}.
    \textit{Condition Problem} causes bugs in most functionalities, especially in {\it Anomaly Detection, Security Protection}, and \textit{Inference}.
\end{tcolorbox} 

\smallskip
\noindent
{\bf Implication:} 
This finding indicates that
(1) it is important to address frequent root causes, i.e., \textit{Logic Error}, \textit{Assignment Problem}, and \textit{Condition Problem}, in various functionalities;
(2) the need for timely updates and meticulous proofreading is urgent to effectively address errors and typos in documentations;
and (3) it is recommended that the maintainers of FL frameworks carefully check the values assigned in the configuration files, especially those related to \textit{Repository Management}.

\subsubsection{Distributions of Fix Patterns across Functionalities}

\begin{figure}[t]
  \centering
  \includegraphics[width=1\linewidth]{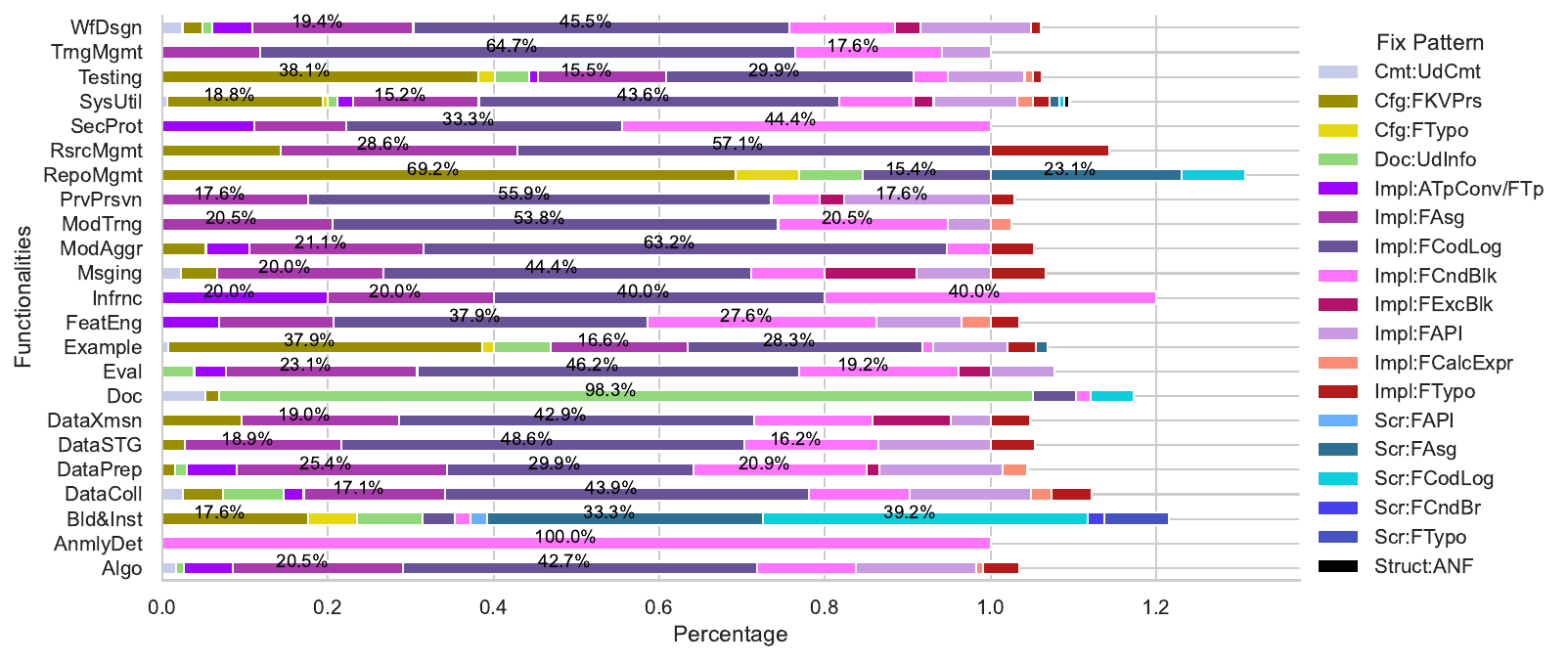}
  \caption{The distributions of fix patterns across functionalities.}
  \label{fig:fix-pattern-on-functionalities}
\end{figure}

Fig.~\ref{fig:fix-pattern-on-functionalities} shows the distributions of fix patterns across functionalities.
We can observe that in most functionalities, bugs are fixed by modifying source code implementation.
However, in \textit{Build~\& Installation},
\textit{Documentation}, 
and \textit{Repository Management}, 
fixings in scripts (e.g., \pr{https://github.com/NVIDIA/NVFlare/pull/920}{920} in NVFlare), documentations (e.g., \pr{https://github.com/OpenMined/PySyft/pull/203}{203} in PySyft), and configuration files (e.g., \pr{https://github.com/OpenMined/PySyft/pull/4928}{4928} in PySyft)
become the most frequent fix patterns, respectively.
Notably, in \textit{Example} and \textit{Testing}, fix patterns in configuration files
(e.g., \pr{https://github.com/FederatedAI/FATE/pull/916/commits/46f641bb39f50936d02cbf18f03ed253b81d8eed}{916}
in FATE)
closely follow those in source code implementation, as the second most frequent category of fix patterns.
In addition, the proportion of fix patterns in scripts is generally negligible in most functionalities, except for \textit{Build~\& Installation} and \textit{Repository Management}.
Specifically, in \textit{Build~\& Installation}, fix patterns in scripts cover 80.4\% (41/51) of bugs.
In \textit{Repository Management}, fixing incorrect assignments in scripts constitutes 23.1\% (3/13) of bugs.

\begin{tcolorbox}[size=title]
    {\bf Finding 9:}
    The majority of bugs are fixed in source code implementation in most functionalities, except for {\it Build~\& Installation, Documentation}, and \textit{Repository Management}.
    The proportion of fix patterns in scripts is generally negligible in most functionalities, except for \textit{Build~\& Installation} and \textit{Repository Management}.
\end{tcolorbox} 

\smallskip
\noindent
{\bf Implication:} 
This finding emphasizes the importance of bug localization and repair in source code implementation in most functionalities. 
In addition, it is crucial to address errors in scripts and configuration files, especially for tools that target bugs in \textit{Build~\& Installation} and \textit{Repository Management}.

\section{Threats to Validity} \label{sec:threats2validity}

{\bfseries Internal threats.}
The major internal threat to validity in our classification and labeling process
is the subjectivity.
To mitigate subjectivity, two authors work together to construct taxonomies by following the open-coding scheme and the methodology adapted by the existing ones~\cite{GarciaF0AXC20, ShenM0TCC21, QuanGXCLL22, JiaZWHL21},
Refinements and adjustments are carried out iteratively whenever conflicts and disagreements are encountered. Furthermore, we leverage Cohen's Kappa coefficient~\cite{cohen1960coefficient} to measure the agreement, which finally demonstrates high agreement between the two authors.

\smallskip
\noindent {\bfseries External threats.}
The external threat to validity mainly lies in the dataset we collect. 
For generalization, we broadly collect both popular FL frameworks and promising yet small ones. 
For accuracy, we select closed issues and merged pull requests that were tagged with ``bug'' by the maintainers. 
Given the prevalence of disorganized pull requests and ambiguous bug reports, we made our best efforts 
to understand bugs, bug symptoms, and their root causes and fixes. 
Ignoring these challenging cases would lead to less convincing
dataset and thus results. 

\balance
\section{Related Work}  \label{sec:related_work}

A number of empirical studies have been conducted on ML/DL frameworks. 
Thung et al.~\cite{ThungWLJ12} analyzed Apache Mahout, Lucene, and OpenNLP 
to understand bug types, frequency and their corresponding fix patterns.
Nejadgholi and Yang~\cite{NejadgholiY19} studied the oracle approximation assertions utilized
in four popular DL libraries (TensorFlow, Theano, PyTorch, and Keras).
Du et al.~\cite{DuXS20} focused on TensorFlow and collected 2,285 bugs to study fault triggering conditions.
%
Very recently, Du et al.~\cite{DuSLA23} extended research targets to three widely-used DL frameworks (TensorFlow, MXNet, and PaddlePaddle) and gathered 3,555 bug reports.
Pham et al.~\cite{PhamQWLRTYN20} was the first to study the variances caused by DL frameworks to models' accuracy and training efficiency. 
%
Sun et al.~\cite{SunZWDBC21} conducted a study with 939 bugs from five ML frameworks on their categories, fix patterns, fixing scale, etc. 
Jia et al.~\cite{JiaZWHL21} only studied TensorFlow to analyze the bug characteristics therein.
They found \textit{Functional Error} and \textit{Crash} were the two most frequent symptoms, similar to our findings on symptoms in FL frameworks.
Yang et al.~\cite{YangHXF22} summarized 1,127 bug reports from eight popular DL frameworks and labeled the bug types, root causes, and symptoms. Also, they explored the fix patterns and provided insights and suggestions to practitioners.
Quan et al.~\cite{QuanGXCLL22} carried out the first empirical study on quality problems of JavaScript-based DL systems by collecting 700 bugs from 72 repositories from GitHub. 
%
Chen et al.~\cite{ChenLSQJL23} conducted a large-scale study of 1,000 bugs from four DL frameworks (TensorFlow, PyTorch, MXNet, and DL4J). 
Also, they measured the test coverage accomplished by three state-of-the-art testing techniques.

All the aforementioned studies exclusively focused on ML/DL frameworks. 
Most of them focused on bugs while the others concentrated on fault triggering conditions~\cite{DuXS20, DuSLA23} and variances~\cite{PhamQWLRTYN20}. 
In comparison, we target bugs in FL frameworks, a more security-critical yet complicated domain. 
At the methodology level, we conduct a cross analysis of distributions on 23 functionalities based on their functions and related tasks, which are different from the five-level architecture~\cite{ChenLSQJL23} and the components identified by repository directories~\cite{JiaZWHL21}.

\section{Conclusion}  \label{sec:conclusion}

In this paper, we conducted the first empirical study on 1,119 bugs from 17 open-source FL frameworks on GitHub by manual collection and analysis. 
We established taxonomies of symptoms, root causes, and fix patterns of the bugs and functionalities where the bugs reside. 
Furthermore, we studied the correlations among symptoms, root causes, and fix patterns, as well as distributions of those categories across functionalities. 
On the basis of the results revealed by the study, we summarized nine findings, discussed the implications, and provided suggestions as guidance to prevent and mitigate those bugs for FL frameworks developers and researchers.

\section*{Data-Availability Statement}
The resulting dataset from our study with detailed sub-categories can be found at~\cite{FLBugdataset}, to replicate or reproduce our results, and allow other researchers and practitioners to build upon our work.  




\begin{thebibliography}{83}
	
	
	\ifx \showCODEN    \undefined \def \showCODEN     #1{\unskip}     \fi
	\ifx \showDOI      \undefined \def \showDOI       #1{#1}\fi
	\ifx \showISBNx    \undefined \def \showISBNx     #1{\unskip}     \fi
	\ifx \showISBNxiii \undefined \def \showISBNxiii  #1{\unskip}     \fi
	\ifx \showISSN     \undefined \def \showISSN      #1{\unskip}     \fi
	\ifx \showLCCN     \undefined \def \showLCCN      #1{\unskip}     \fi
	\ifx \shownote     \undefined \def \shownote      #1{#1}          \fi
	\ifx \showarticletitle \undefined \def \showarticletitle #1{#1}   \fi
	\ifx \showURL      \undefined \def \showURL       {\relax}        \fi
	\providecommand\bibfield[2]{#2}
	\providecommand\bibinfo[2]{#2}
	\providecommand\natexlab[1]{#1}
	\providecommand\showeprint[2][]{arXiv:#2}
	
	\bibitem[chi(2021)]%
	{china_data_security_law}
	\bibinfo{year}{2021}\natexlab{}.
	\newblock \bibinfo{title}{Data Security Law of the People's Republic of China}.
	\newblock
	\newblock
	
	
	\bibitem[FLB(2023)]%
	{FLBugdataset}
	\bibinfo{year}{2023}\natexlab{}.
	\newblock \bibinfo{booktitle}{\emph{Our Dataset for the Study of FL bugs}}.
	\newblock
	\urldef\tempurl%
	\url{https://anonymous.4open.science/r/fl-framework-bug-data-543C}
	\showURL{%
		\tempurl}
	
	
	\bibitem[Beck et~al\mbox{.}(2001)]%
	{beck2001manifesto}
	\bibfield{author}{\bibinfo{person}{Kent Beck}, \bibinfo{person}{Mike Beedle},
		\bibinfo{person}{Arie Van~Bennekum}, \bibinfo{person}{Alistair Cockburn},
		\bibinfo{person}{Ward Cunningham}, \bibinfo{person}{Martin Fowler},
		\bibinfo{person}{James Grenning}, \bibinfo{person}{Jim Highsmith},
		\bibinfo{person}{Andrew Hunt}, \bibinfo{person}{Ron Jeffries},
		{et~al\mbox{.}}} \bibinfo{year}{2001}\natexlab{}.
	\newblock \showarticletitle{Manifesto for agile software development}.
	\newblock  (\bibinfo{year}{2001}).
	\newblock
	
	
	\bibitem[Beutel et~al\mbox{.}(2020)]%
	{beutel2020flower}
	\bibfield{author}{\bibinfo{person}{Daniel~J. Beutel}, \bibinfo{person}{Taner
			Topal}, \bibinfo{person}{Akhil Mathur}, \bibinfo{person}{Xinchi Qiu},
		\bibinfo{person}{Titouan Parcollet}, {and} \bibinfo{person}{Nicholas~D.
			Lane}.} \bibinfo{year}{2020}\natexlab{}.
	\newblock \showarticletitle{Flower: {A} Friendly Federated Learning Research
		Framework}.
	\newblock \bibinfo{journal}{\emph{CoRR}}  \bibinfo{volume}{abs/2007.14390}
	(\bibinfo{year}{2020}).
	\newblock
	
	
	\bibitem[Bonawitz et~al\mbox{.}(2019)]%
	{BonawitzEGHIIKK19}
	\bibfield{author}{\bibinfo{person}{Kallista~A. Bonawitz},
		\bibinfo{person}{Hubert Eichner}, \bibinfo{person}{Wolfgang Grieskamp},
		\bibinfo{person}{Dzmitry Huba}, \bibinfo{person}{Alex Ingerman},
		\bibinfo{person}{Vladimir Ivanov}, \bibinfo{person}{Chlo{\'{e}} Kiddon},
		\bibinfo{person}{Jakub Kone{\v{c}}n{\'y}}, \bibinfo{person}{Stefano
			Mazzocchi}, \bibinfo{person}{Brendan McMahan}, \bibinfo{person}{Timon~Van
			Overveldt}, \bibinfo{person}{David Petrou}, \bibinfo{person}{Daniel Ramage},
		{and} \bibinfo{person}{Jason Roselander}.} \bibinfo{year}{2019}\natexlab{}.
	\newblock \showarticletitle{Towards Federated Learning at Scale: System
		Design}. In \bibinfo{booktitle}{\emph{Proceedings of Machine Learning and
			Systems 2019, MLSys 2019}}.
	\newblock
	
	
	\bibitem[Bonawitz et~al\mbox{.}(2017)]%
	{BonawitzIKMMPRS17}
	\bibfield{author}{\bibinfo{person}{Kallista~A. Bonawitz},
		\bibinfo{person}{Vladimir Ivanov}, \bibinfo{person}{Ben Kreuter},
		\bibinfo{person}{Antonio Marcedone}, \bibinfo{person}{H.~Brendan McMahan},
		\bibinfo{person}{Sarvar Patel}, \bibinfo{person}{Daniel Ramage},
		\bibinfo{person}{Aaron Segal}, {and} \bibinfo{person}{Karn Seth}.}
	\bibinfo{year}{2017}\natexlab{}.
	\newblock \showarticletitle{Practical Secure Aggregation for Privacy-Preserving
		Machine Learning}. In \bibinfo{booktitle}{\emph{Proceedings of the 2017 {ACM}
			{SIGSAC} Conference on Computer and Communications Security, {CCS} 2017}}.
	\newblock
	
	
	\bibitem[ByteDance(2022)]%
	{Fedlearner}
	\bibfield{author}{\bibinfo{person}{ByteDance}.}
	\bibinfo{year}{2022}\natexlab{}.
	\newblock \bibinfo{booktitle}{\emph{{Fedlearner}}}.
	\newblock
	\urldef\tempurl%
	\url{https://github.com/bytedance/fedlearner}
	\showURL{%
		\tempurl}
	
	
	\bibitem[Cacho et~al\mbox{.}(2014)]%
	{CachoBAPGCSCFG14}
	\bibfield{author}{\bibinfo{person}{N{\'{e}}lio Cacho},
		\bibinfo{person}{Eiji~Adachi Barbosa}, \bibinfo{person}{Juliana Araujo},
		\bibinfo{person}{Frederico Pranto}, \bibinfo{person}{Alessandro~F. Garcia},
		\bibinfo{person}{Thiago C{\'{e}}sar}, \bibinfo{person}{Eliezio Soares},
		\bibinfo{person}{Arthur Cassio}, \bibinfo{person}{Thomas Filipe}, {and}
		\bibinfo{person}{Israel Garc{\'{\i}}a}.} \bibinfo{year}{2014}\natexlab{}.
	\newblock \showarticletitle{How Does Exception Handling Behavior Evolve? An
		Exploratory Study in Java and C{\#} Applications}. In
	\bibinfo{booktitle}{\emph{Proceedings of the 30th {IEEE} International
			Conference on Software Maintenance and Evolution (ICSME)}}.
	\bibinfo{pages}{31--40}.
	\newblock
	
	
	\bibitem[Chen et~al\mbox{.}(2023)]%
	{ChenLSQJL23}
	\bibfield{author}{\bibinfo{person}{Junjie Chen}, \bibinfo{person}{Yihua Liang},
		\bibinfo{person}{Qingchao Shen}, \bibinfo{person}{Jiajun Jiang}, {and}
		\bibinfo{person}{Shuochuan Li}.} \bibinfo{year}{2023}\natexlab{}.
	\newblock \showarticletitle{Toward Understanding Deep Learning Framework Bugs}.
	\newblock \bibinfo{journal}{\emph{ACM Trans. Softw. Eng. Methodol.}}
	(\bibinfo{year}{2023}).
	\newblock
	
	
	\bibitem[Coelho et~al\mbox{.}(2015)]%
	{CoelhoAGD15}
	\bibfield{author}{\bibinfo{person}{Roberta Coelho}, \bibinfo{person}{Lucas
			Almeida}, \bibinfo{person}{Georgios Gousios}, {and} \bibinfo{person}{Arie van
			Deursen}.} \bibinfo{year}{2015}\natexlab{}.
	\newblock \showarticletitle{Unveiling Exception Handling Bug Hazards in Android
		Based on GitHub and Google Code Issues}. In
	\bibinfo{booktitle}{\emph{Proceedings of the 12th {IEEE/ACM} Working
			Conference on Mining Software Repositories ({MSR})}}.
	\bibinfo{pages}{134--145}.
	\newblock
	
	
	\bibitem[Cohen(1960)]%
	{cohen1960coefficient}
	\bibfield{author}{\bibinfo{person}{Jacob Cohen}.}
	\bibinfo{year}{1960}\natexlab{}.
	\newblock \showarticletitle{A coefficient of agreement for nominal scales}.
	\newblock \bibinfo{journal}{\emph{Educational and psychological measurement}}
	\bibinfo{volume}{20}, \bibinfo{number}{1} (\bibinfo{year}{1960}),
	\bibinfo{pages}{37--46}.
	\newblock
	
	
	\bibitem[Daga et~al\mbox{.}(2023)]%
	{flame}
	\bibfield{author}{\bibinfo{person}{Harshit Daga}, \bibinfo{person}{Jaemin
			Shin}, \bibinfo{person}{Dhruv Garg}, \bibinfo{person}{Ada Gavrilovska},
		\bibinfo{person}{Myungjin Lee}, {and} \bibinfo{person}{Ramana~Rao Kompella}.}
	\bibinfo{year}{2023}\natexlab{}.
	\newblock \showarticletitle{Federated Learning Operations Made Simple with
		Flame}.
	\newblock \bibinfo{journal}{\emph{CoRR}}  \bibinfo{volume}{abs/2305.05118}
	(\bibinfo{year}{2023}).
	\newblock
	
	
	\bibitem[Dimitriadis et~al\mbox{.}(2022)]%
	{FLUTE}
	\bibfield{author}{\bibinfo{person}{Dimitrios Dimitriadis},
		\bibinfo{person}{Mirian~Hipolito Garcia}, \bibinfo{person}{Daniel~Madrigal
			Diaz}, \bibinfo{person}{Andre Manoel}, {and} \bibinfo{person}{Robert Sim}.}
	\bibinfo{year}{2022}\natexlab{}.
	\newblock \showarticletitle{{FLUTE:} {A} Scalable, Extensible Framework for
		High-Performance Federated Learning Simulations}.
	\newblock \bibinfo{journal}{\emph{CoRR}}  \bibinfo{volume}{abs/2203.13789}
	(\bibinfo{year}{2022}).
	\newblock
	
	
	\bibitem[Du et~al\mbox{.}(2023)]%
	{DuSLA23}
	\bibfield{author}{\bibinfo{person}{Xiaoting Du}, \bibinfo{person}{Yulei Sui},
		\bibinfo{person}{Zhihao Liu}, {and} \bibinfo{person}{Jun Ai}.}
	\bibinfo{year}{2023}\natexlab{}.
	\newblock \showarticletitle{An Empirical Study of Fault Triggers in Deep
		Learning Frameworks}.
	\newblock \bibinfo{journal}{\emph{{IEEE} Trans. Dependable Secur. Comput.}}
	\bibinfo{volume}{20}, \bibinfo{number}{4} (\bibinfo{year}{2023}),
	\bibinfo{pages}{2696--2712}.
	\newblock
	
	
	\bibitem[Du et~al\mbox{.}(2020)]%
	{DuXS20}
	\bibfield{author}{\bibinfo{person}{Xiaoting Du}, \bibinfo{person}{Guanping
			Xiao}, {and} \bibinfo{person}{Yulei Sui}.} \bibinfo{year}{2020}\natexlab{}.
	\newblock \showarticletitle{Fault Triggers in the TensorFlow Framework: An
		Experience Report}. In \bibinfo{booktitle}{\emph{31st {IEEE} International
			Symposium on Software Reliability Engineering, {ISSRE} 2020}}.
	\newblock
	
	
	\bibitem[Ecosystem(2021)]%
	{FATEConfSys}
	\bibfield{author}{\bibinfo{person}{Federated~AI Ecosystem}.}
	\bibinfo{year}{2021}\natexlab{}.
	\newblock \bibinfo{booktitle}{\emph{FATE}}.
	\newblock
	\urldef\tempurl%
	\url{https://github.com/FederatedAI/FATE/pull/3391}
	\showURL{%
		\tempurl}
	
	
	\bibitem[Ekmefjord et~al\mbox{.}({[n.\,d.]})]%
	{EkmefjordAAASST22}
	\bibfield{author}{\bibinfo{person}{Morgan Ekmefjord}, \bibinfo{person}{Addi
			Ait{-}Mlouk}, \bibinfo{person}{Sadi Alawadi}, \bibinfo{person}{Mattias
			{\AA}kesson}, \bibinfo{person}{Prashant Singh}, \bibinfo{person}{Ola Spjuth},
		\bibinfo{person}{Salman Toor}, {and} \bibinfo{person}{Andreas Hellander}.}
	\bibinfo{year}{[n.\,d.]}\natexlab{}.
	\newblock \showarticletitle{Scalable federated machine learning with FEDn}. In
	\bibinfo{booktitle}{\emph{22nd {IEEE} International Symposium on Cluster,
			Cloud and Internet Computing, CCGrid 2022}}.
	\newblock
	
	
	\bibitem[{European Commission}(2016)]%
	{european_commission_regulation_2016}
	\bibfield{author}{\bibinfo{person}{{European Commission}}.}
	\bibinfo{year}{2016}\natexlab{}.
	\newblock \bibinfo{title}{Regulation ({EU}) 2016/679 of the {European}
		{Parliament} and of the {Council} on the protection of natural persons with
		regard to the processing of personal data and on the free movement of such
		data, and repealing {Directive} 95/46/{EC} ({General} {Data} {Protection}
		{Regulation}) ({Text} with {EEA} relevance)}.
	\newblock
	\newblock
	
	
	\bibitem[Fang et~al\mbox{.}(2020)]%
	{FangCJG20}
	\bibfield{author}{\bibinfo{person}{Minghong Fang}, \bibinfo{person}{Xiaoyu
			Cao}, \bibinfo{person}{Jinyuan Jia}, {and} \bibinfo{person}{Neil~Zhenqiang
			Gong}.} \bibinfo{year}{2020}\natexlab{}.
	\newblock \showarticletitle{Local Model Poisoning Attacks to Byzantine-Robust
		Federated Learning}. In \bibinfo{booktitle}{\emph{29th {USENIX} Security
			Symposium, {USENIX} Security 2020}},
	\bibfield{editor}{\bibinfo{person}{Srdjan Capkun} {and}
		\bibinfo{person}{Franziska Roesner}} (Eds.). \bibinfo{publisher}{{USENIX}
		Association}, \bibinfo{pages}{1605--1622}.
	\newblock
	
	
	\bibitem[Franco et~al\mbox{.}(2017)]%
	{FrancoGR17}
	\bibfield{author}{\bibinfo{person}{Anthony~Di Franco}, \bibinfo{person}{Hui
			Guo}, {and} \bibinfo{person}{Cindy Rubio{-}Gonz{\'{a}}lez}.}
	\bibinfo{year}{2017}\natexlab{}.
	\newblock \showarticletitle{A comprehensive study of real-world numerical bug
		characteristics}. In \bibinfo{booktitle}{\emph{Proceedings of the 32nd
			{IEEE/ACM} International Conference on Automated Software Engineering
			({ASE}}}. \bibinfo{pages}{509--519}.
	\newblock
	
	
	\bibitem[Fung et~al\mbox{.}(2018)]%
	{FungYB18}
	\bibfield{author}{\bibinfo{person}{Clement Fung}, \bibinfo{person}{Chris J.~M.
			Yoon}, {and} \bibinfo{person}{Ivan Beschastnikh}.}
	\bibinfo{year}{2018}\natexlab{}.
	\newblock \showarticletitle{Mitigating Sybils in Federated Learning Poisoning}.
	\newblock \bibinfo{journal}{\emph{CoRR}}  \bibinfo{volume}{abs/1808.04866}
	(\bibinfo{year}{2018}).
	\newblock
	
	
	\bibitem[Galtier and Marini(2019)]%
	{Substra}
	\bibfield{author}{\bibinfo{person}{Mathieu~N. Galtier} {and}
		\bibinfo{person}{Camille Marini}.} \bibinfo{year}{2019}\natexlab{}.
	\newblock \showarticletitle{Substra: a framework for privacy-preserving,
		traceable and collaborative Machine Learning}.
	\newblock \bibinfo{journal}{\emph{CoRR}}  \bibinfo{volume}{abs/1910.11567}
	(\bibinfo{year}{2019}).
	\newblock
	
	
	\bibitem[Garcia et~al\mbox{.}(2020)]%
	{GarciaF0AXC20}
	\bibfield{author}{\bibinfo{person}{Joshua Garcia}, \bibinfo{person}{Yang Feng},
		\bibinfo{person}{Junjie Shen}, \bibinfo{person}{Sumaya Almanee},
		\bibinfo{person}{Yuan Xia}, {and} \bibinfo{person}{Qi~Alfred Chen}.}
	\bibinfo{year}{2020}\natexlab{}.
	\newblock \showarticletitle{A comprehensive study of autonomous vehicle bugs}.
	In \bibinfo{booktitle}{\emph{Proceedings of the 42nd International Conference
			on Software Engineering ({ICSE})}}. \bibinfo{pages}{385--396}.
	\newblock
	
	
	\bibitem[Geyer et~al\mbox{.}(2017)]%
	{GeyerKN17}
	\bibfield{author}{\bibinfo{person}{Robin~C. Geyer}, \bibinfo{person}{Tassilo
			Klein}, {and} \bibinfo{person}{Moin Nabi}.} \bibinfo{year}{2017}\natexlab{}.
	\newblock \showarticletitle{Differentially Private Federated Learning: {A}
		Client Level Perspective}.
	\newblock \bibinfo{journal}{\emph{CoRR}}  \bibinfo{volume}{abs/1712.07557}
	(\bibinfo{year}{2017}).
	\newblock
	
	
	\bibitem[Han et~al\mbox{.}(2011)]%
	{HanKP2011}
	\bibfield{author}{\bibinfo{person}{Jiawei Han}, \bibinfo{person}{Micheline
			Kamber}, {and} \bibinfo{person}{Jian Pei}.} \bibinfo{year}{2011}\natexlab{}.
	\newblock \bibinfo{booktitle}{\emph{Data Mining: Concepts and Techniques, 3rd
			edition}}.
	\newblock \bibinfo{publisher}{Morgan Kaufmann}.
	\newblock
	
	
	\bibitem[He et~al\mbox{.}(2020)]%
	{he2020fedml}
	\bibfield{author}{\bibinfo{person}{Chaoyang He}, \bibinfo{person}{Songze Li},
		\bibinfo{person}{Jinhyun So}, \bibinfo{person}{Mi Zhang},
		\bibinfo{person}{Hongyi Wang}, \bibinfo{person}{Xiaoyang Wang},
		\bibinfo{person}{Praneeth Vepakomma}, \bibinfo{person}{Abhishek Singh},
		\bibinfo{person}{Hang Qiu}, \bibinfo{person}{Li Shen},
		\bibinfo{person}{Peilin Zhao}, \bibinfo{person}{Yan Kang},
		\bibinfo{person}{Yang Liu}, \bibinfo{person}{Ramesh Raskar},
		\bibinfo{person}{Qiang Yang}, \bibinfo{person}{Murali Annavaram}, {and}
		\bibinfo{person}{Salman Avestimehr}.} \bibinfo{year}{2020}\natexlab{}.
	\newblock \showarticletitle{FedML: {A} Research Library and Benchmark for
		Federated Machine Learning}.
	\newblock \bibinfo{journal}{\emph{CoRR}}  \bibinfo{volume}{abs/2007.13518}
	(\bibinfo{year}{2020}).
	\newblock
	
	
	\bibitem[Hu et~al\mbox{.}(2020)]%
	{GFL}
	\bibfield{author}{\bibinfo{person}{Yifan Hu}, \bibinfo{person}{Wei Xia},
		\bibinfo{person}{Jun Xiao}, {and} \bibinfo{person}{Chao Wu}.}
	\bibinfo{year}{2020}\natexlab{}.
	\newblock \showarticletitle{{GFL:} {A} Decentralized Federated Learning
		Framework Based On Blockchain}.
	\newblock \bibinfo{journal}{\emph{CoRR}}  \bibinfo{volume}{abs/2010.10996}
	(\bibinfo{year}{2020}).
	\newblock
	
	
	\bibitem[Huawei Technologies~Co.(2023)]%
	{mindspore23}
	\bibfield{author}{\bibinfo{person}{Ltd. Huawei Technologies~Co.}}
	\bibinfo{year}{2023}\natexlab{}.
	\newblock \bibinfo{booktitle}{\emph{Huawei MindSpore AI Development
			Framework}}.
	\newblock \bibinfo{publisher}{Springer Nature Singapore},
	\bibinfo{address}{Singapore}, \bibinfo{pages}{137--162}.
	\newblock
	
	
	\bibitem[iFLYTEK(2022)]%
	{iFLearner}
	\bibfield{author}{\bibinfo{person}{iFLYTEK}.} \bibinfo{year}{2022}\natexlab{}.
	\newblock \bibinfo{booktitle}{\emph{iFLearner}}.
	\newblock
	\urldef\tempurl%
	\url{https://github.com/iflytek/iflearner}
	\showURL{%
		\tempurl}
	
	
	\bibitem[Islam et~al\mbox{.}(2019)]%
	{IslamNPR19}
	\bibfield{author}{\bibinfo{person}{Md~Johirul Islam}, \bibinfo{person}{Giang
			Nguyen}, \bibinfo{person}{Rangeet Pan}, {and} \bibinfo{person}{Hridesh
			Rajan}.} \bibinfo{year}{2019}\natexlab{}.
	\newblock \showarticletitle{A comprehensive study on deep learning bug
		characteristics}. In \bibinfo{booktitle}{\emph{Proceedings of the {ACM} Joint
			Meeting on European Software EngineeringConference and Symposium on the
			Foundations of Software Engineering ({ESEC/SIGSOFT})}}.
	\newblock
	
	
	\bibitem[Jia et~al\mbox{.}(2021)]%
	{JiaZWHL21}
	\bibfield{author}{\bibinfo{person}{Li Jia}, \bibinfo{person}{Hao Zhong},
		\bibinfo{person}{Xiaoyin Wang}, \bibinfo{person}{Linpeng Huang}, {and}
		\bibinfo{person}{Xuansheng Lu}.} \bibinfo{year}{2021}\natexlab{}.
	\newblock \showarticletitle{The symptoms, causes, and repairs of bugs inside a
		deep learning library}.
	\newblock \bibinfo{journal}{\emph{J. Syst. Softw.}}  \bibinfo{volume}{177}
	(\bibinfo{year}{2021}), \bibinfo{pages}{110935}.
	\newblock
	
	
	\bibitem[Jin et~al\mbox{.}(2012)]%
	{JinSSSL12}
	\bibfield{author}{\bibinfo{person}{Guoliang Jin}, \bibinfo{person}{Linhai
			Song}, \bibinfo{person}{Xiaoming Shi}, \bibinfo{person}{Joel Scherpelz},
		{and} \bibinfo{person}{Shan Lu}.} \bibinfo{year}{2012}\natexlab{}.
	\newblock \showarticletitle{Understanding and detecting real-world performance
		bugs}. In \bibinfo{booktitle}{\emph{Proceedings of the {ACM} {SIGPLAN}
			Conference on Programming Language Design and Implementation ( {PLDI})}}.
	\bibinfo{pages}{77--88}.
	\newblock
	
	
	\bibitem[Kone{\v{c}}n{\'y} et~al\mbox{.}(2015)]%
	{KonecnyMR15}
	\bibfield{author}{\bibinfo{person}{Jakub Kone{\v{c}}n{\'y}},
		\bibinfo{person}{Brendan McMahan}, {and} \bibinfo{person}{Daniel Ramage}.}
	\bibinfo{year}{2015}\natexlab{}.
	\newblock \showarticletitle{Federated Optimization: Distributed Optimization
		Beyond the Datacenter}.
	\newblock \bibinfo{journal}{\emph{CoRR}}  \bibinfo{volume}{abs/1511.03575}
	(\bibinfo{year}{2015}).
	\newblock
	
	
	\bibitem[Koyuncu et~al\mbox{.}(2018)]%
	{Koyuncu2018FixMinerMR}
	\bibfield{author}{\bibinfo{person}{Anil Koyuncu}, \bibinfo{person}{Kui Liu},
		\bibinfo{person}{Tegawend{\'e}~F. Bissyand{\'e}}, \bibinfo{person}{Dongsun
			Kim}, \bibinfo{person}{Jacques Klein}, \bibinfo{person}{Monperrus Martin},
		{and} \bibinfo{person}{Yves~Le Traon}.} \bibinfo{year}{2018}\natexlab{}.
	\newblock \showarticletitle{FixMiner: Mining relevant fix patterns for
		automated program repair}.
	\newblock \bibinfo{journal}{\emph{Empirical Software Engineering}}
	(\bibinfo{year}{2018}).
	\newblock
	
	
	\bibitem[Labs(2022)]%
	{Cfg_FKVPrs}
	\bibfield{author}{\bibinfo{person}{Flower Labs}.}
	\bibinfo{year}{2022}\natexlab{}.
	\newblock \bibinfo{booktitle}{\emph{Flower}}.
	\newblock
	\urldef\tempurl%
	\url{https://github.com/adap/flower/pull/1344}
	\showURL{%
		\tempurl}
	
	
	\bibitem[Lai et~al\mbox{.}(2022)]%
	{fedscale-icml22}
	\bibfield{author}{\bibinfo{person}{Fan Lai}, \bibinfo{person}{Yinwei Dai},
		\bibinfo{person}{Sanjay~S. Singapuram}, \bibinfo{person}{Jiachen Liu},
		\bibinfo{person}{Xiangfeng Zhu}, \bibinfo{person}{Harsha~V. Madhyastha},
		{and} \bibinfo{person}{Mosharaf Chowdhury}.} \bibinfo{year}{2022}\natexlab{}.
	\newblock \showarticletitle{{FedScale}: Benchmarking Model and System
		Performance of Federated Learning at Scale}. In
	\bibinfo{booktitle}{\emph{International Conference on Machine Learning,
			{ICML} 2022}}.
	\newblock
	
	
	\bibitem[Leesatapornwongsa et~al\mbox{.}(2016)]%
	{Leesatapornwongsa16}
	\bibfield{author}{\bibinfo{person}{Tanakorn Leesatapornwongsa},
		\bibinfo{person}{Jeffrey~F. Lukman}, \bibinfo{person}{Shan Lu}, {and}
		\bibinfo{person}{Haryadi~S. Gunawi}.} \bibinfo{year}{2016}\natexlab{}.
	\newblock \showarticletitle{TaxDC: {A} Taxonomy of Non-Deterministic
		Concurrency Bugs in Datacenter Distributed Systems}. In
	\bibinfo{booktitle}{\emph{Proceedings of the Twenty-First International
			Conference on Architectural Support for Programming Languages and Operating
			Systems ({ASPLOS})}}. \bibinfo{pages}{517--530}.
	\newblock
	
	
	\bibitem[Li et~al\mbox{.}(2023)]%
	{FedTree}
	\bibfield{author}{\bibinfo{person}{Qinbin Li}, \bibinfo{person}{Zhaomin Wu},
		\bibinfo{person}{Yanzheng Cai}, \bibinfo{person}{Yuxuan Han},
		\bibinfo{person}{Ching~Man Yung}, \bibinfo{person}{Tianyuan Fu}, {and}
		\bibinfo{person}{Bingsheng He}.} \bibinfo{year}{2023}\natexlab{}.
	\newblock \showarticletitle{FedTree: A Federated Learning System For Trees}. In
	\bibinfo{booktitle}{\emph{Proceedings of Machine Learning and Systems}}.
	\newblock
	
	
	\bibitem[Li et~al\mbox{.}(2020)]%
	{LiSM20}
	\bibfield{author}{\bibinfo{person}{Zengpeng Li}, \bibinfo{person}{Vishal
			Sharma}, {and} \bibinfo{person}{Saraju~P. Mohanty}.}
	\bibinfo{year}{2020}\natexlab{}.
	\newblock \showarticletitle{Preserving Data Privacy via Federated Learning:
		Challenges and Solutions}.
	\newblock \bibinfo{journal}{\emph{{IEEE} Consumer Electron. Mag.}}
	\bibinfo{volume}{9}, \bibinfo{number}{3} (\bibinfo{year}{2020}),
	\bibinfo{pages}{8--16}.
	\newblock
	
	
	\bibitem[Liu et~al\mbox{.}(2021)]%
	{LiuFCXY21}
	\bibfield{author}{\bibinfo{person}{Yang Liu}, \bibinfo{person}{Tao Fan},
		\bibinfo{person}{Tianjian Chen}, \bibinfo{person}{Qian Xu}, {and}
		\bibinfo{person}{Qiang Yang}.} \bibinfo{year}{2021}\natexlab{}.
	\newblock \showarticletitle{{FATE:} An Industrial Grade Platform for
		Collaborative Learning With Data Protection}.
	\newblock \bibinfo{journal}{\emph{J. Mach. Learn. Res.}}  \bibinfo{volume}{22}
	(\bibinfo{year}{2021}), \bibinfo{pages}{226:1--226:6}.
	\newblock
	
	
	\bibitem[Lo et~al\mbox{.}(2022)]%
	{LoLWPZ21}
	\bibfield{author}{\bibinfo{person}{Sin~Kit Lo}, \bibinfo{person}{Qinghua Lu},
		\bibinfo{person}{Chen Wang}, \bibinfo{person}{Hye{-}Young Paik}, {and}
		\bibinfo{person}{Liming Zhu}.} \bibinfo{year}{2022}\natexlab{}.
	\newblock \showarticletitle{A Systematic Literature Review on Federated Machine
		Learning: From a Software Engineering Perspective}.
	\newblock \bibinfo{journal}{\emph{{ACM} Comput. Surv.}} \bibinfo{volume}{54},
	\bibinfo{number}{5} (\bibinfo{year}{2022}), \bibinfo{pages}{95:1--95:39}.
	\newblock
	
	
	\bibitem[Lu et~al\mbox{.}(2008)]%
	{LuPSZ08}
	\bibfield{author}{\bibinfo{person}{Shan Lu}, \bibinfo{person}{Soyeon Park},
		\bibinfo{person}{Eunsoo Seo}, {and} \bibinfo{person}{Yuanyuan Zhou}.}
	\bibinfo{year}{2008}\natexlab{}.
	\newblock \showarticletitle{Learning from mistakes: a comprehensive study on
		real world concurrency bug characteristics}. In
	\bibinfo{booktitle}{\emph{Proceedings of the 13th International Conference on
			Architectural Support for Programming Languages and Operating Systems
			({ASPLOS})}}. \bibinfo{pages}{329--339}.
	\newblock
	
	
	\bibitem[Ludwig et~al\mbox{.}(2020)]%
	{ibm-federated-learning}
	\bibfield{author}{\bibinfo{person}{Heiko Ludwig}, \bibinfo{person}{Nathalie
			Baracaldo}, \bibinfo{person}{Gegi Thomas}, \bibinfo{person}{Yi Zhou},
		\bibinfo{person}{Ali Anwar}, \bibinfo{person}{Shashank Rajamoni},
		\bibinfo{person}{Yuya~Jeremy Ong}, \bibinfo{person}{Jayaram Radhakrishnan},
		\bibinfo{person}{Ashish Verma}, \bibinfo{person}{Mathieu Sinn},
		\bibinfo{person}{Mark Purcell}, \bibinfo{person}{Ambrish Rawat},
		\bibinfo{person}{Tran~Ngoc Minh}, \bibinfo{person}{Naoise Holohan},
		\bibinfo{person}{Supriyo Chakraborty}, \bibinfo{person}{Shalisha
			Witherspoon}, \bibinfo{person}{Dean Steuer}, \bibinfo{person}{Laura Wynter},
		\bibinfo{person}{Hifaz Hassan}, \bibinfo{person}{Sean Laguna},
		\bibinfo{person}{Mikhail Yurochkin}, \bibinfo{person}{Mayank Agarwal},
		\bibinfo{person}{Ebube Chuba}, {and} \bibinfo{person}{Annie Abay}.}
	\bibinfo{year}{2020}\natexlab{}.
	\newblock \showarticletitle{{IBM} Federated Learning: an Enterprise Framework
		White Paper {V0.1}}.
	\newblock \bibinfo{journal}{\emph{CoRR}}  \bibinfo{volume}{abs/2007.10987}
	(\bibinfo{year}{2020}).
	\newblock
	
	
	\bibitem[Microsoft(2021)]%
	{fl-simulation}
	\bibfield{author}{\bibinfo{person}{Microsoft}.}
	\bibinfo{year}{2021}\natexlab{}.
	\newblock \bibinfo{booktitle}{\emph{Federated Learning Simulation Framework
			(fl-simulation)}}.
	\newblock
	\urldef\tempurl%
	\url{https://github.com/microsoft/fl-simulation}
	\showURL{%
		\tempurl}
	
	
	\bibitem[Mohassel and Zhang(2017)]%
	{MohasselZ17}
	\bibfield{author}{\bibinfo{person}{Payman Mohassel} {and}
		\bibinfo{person}{Yupeng Zhang}.} \bibinfo{year}{2017}\natexlab{}.
	\newblock \showarticletitle{SecureML: {A} System for Scalable
		Privacy-Preserving Machine Learning}. In \bibinfo{booktitle}{\emph{2017
			{IEEE} Symposium on Security and Privacy, {SP} 2017}}.
	\newblock
	
	
	\bibitem[Nejadgholi and Yang(2019)]%
	{NejadgholiY19}
	\bibfield{author}{\bibinfo{person}{Mahdi Nejadgholi} {and}
		\bibinfo{person}{Jinqiu Yang}.} \bibinfo{year}{2019}\natexlab{}.
	\newblock \showarticletitle{A Study of Oracle Approximations in Testing Deep
		Learning Libraries}. In \bibinfo{booktitle}{\emph{34th {IEEE/ACM}
			International Conference on Automated Software Engineering, {ASE} 2019}}.
	\newblock
	
	
	\bibitem[OpenMined(2020)]%
	{cnd_missing}
	\bibfield{author}{\bibinfo{person}{OpenMined}.}
	\bibinfo{year}{2020}\natexlab{}.
	\newblock \bibinfo{booktitle}{\emph{PySyft}}.
	\newblock
	\urldef\tempurl%
	\url{https://github.com/OpenMined/PySyft/issues/3435}
	\showURL{%
		\tempurl}
	
	
	\bibitem[PaddlePaddle(2022)]%
	{PaddleFL}
	\bibfield{author}{\bibinfo{person}{PaddlePaddle}.}
	\bibinfo{year}{2022}\natexlab{}.
	\newblock \bibinfo{booktitle}{\emph{{PaddleFL}}}.
	\newblock
	\urldef\tempurl%
	\url{https://github.com/PaddlePaddle/PaddleFL}
	\showURL{%
		\tempurl}
	
	
	\bibitem[paritybit.ai(2022)]%
	{XFL}
	\bibfield{author}{\bibinfo{person}{paritybit.ai}.}
	\bibinfo{year}{2022}\natexlab{}.
	\newblock \bibinfo{booktitle}{\emph{XFL}}.
	\newblock
	\urldef\tempurl%
	\url{https://github.com/paritybit-ai/XFL}
	\showURL{%
		\tempurl}
	
	
	\bibitem[Pham et~al\mbox{.}(2020)]%
	{PhamQWLRTYN20}
	\bibfield{author}{\bibinfo{person}{Hung~Viet Pham}, \bibinfo{person}{Shangshu
			Qian}, \bibinfo{person}{Jiannan Wang}, \bibinfo{person}{Thibaud Lutellier},
		\bibinfo{person}{Jonathan Rosenthal}, \bibinfo{person}{Lin Tan},
		\bibinfo{person}{Yaoliang Yu}, {and} \bibinfo{person}{Nachiappan Nagappan}.}
	\bibinfo{year}{2020}\natexlab{}.
	\newblock \showarticletitle{Problems and Opportunities in Training Deep
		Learning Software Systems: An Analysis of Variance}. In
	\bibinfo{booktitle}{\emph{35th {IEEE/ACM} International Conference on
			Automated Software Engineering, {ASE} 2020}}.
	\newblock
	
	
	\bibitem[Phong et~al\mbox{.}(2018)]%
	{PhongAHWM18}
	\bibfield{author}{\bibinfo{person}{Le~Trieu Phong}, \bibinfo{person}{Yoshinori
			Aono}, \bibinfo{person}{Takuya Hayashi}, \bibinfo{person}{Lihua Wang}, {and}
		\bibinfo{person}{Shiho Moriai}.} \bibinfo{year}{2018}\natexlab{}.
	\newblock \showarticletitle{Privacy-Preserving Deep Learning via Additively
		Homomorphic Encryption}.
	\newblock \bibinfo{journal}{\emph{{IEEE} Trans. Inf. Forensics Secur.}}
	\bibinfo{volume}{13}, \bibinfo{number}{5} (\bibinfo{year}{2018}),
	\bibinfo{pages}{1333--1345}.
	\newblock
	
	
	\bibitem[PrimiHub(2022)]%
	{PrimiHub}
	\bibfield{author}{\bibinfo{person}{PrimiHub}.} \bibinfo{year}{2022}\natexlab{}.
	\newblock \bibinfo{booktitle}{\emph{PrimiHub}}.
	\newblock
	\urldef\tempurl%
	\url{https://github.com/primihub/primihub}
	\showURL{%
		\tempurl}
	
	
	\bibitem[Quan et~al\mbox{.}(2022)]%
	{QuanGXCLL22}
	\bibfield{author}{\bibinfo{person}{Lili Quan}, \bibinfo{person}{Qianyu Guo},
		\bibinfo{person}{Xiaofei Xie}, \bibinfo{person}{Sen Chen},
		\bibinfo{person}{Xiaohong Li}, {and} \bibinfo{person}{Yang Liu}.}
	\bibinfo{year}{2022}\natexlab{}.
	\newblock \showarticletitle{Towards Understanding the Faults of
		JavaScript-Based Deep Learning Systems}. In \bibinfo{booktitle}{\emph{37th
			{IEEE/ACM} International Conference on Automated Software Engineering, {ASE}
			2022}}.
	\newblock
	
	
	\bibitem[ray project(2022)]%
	{RayFed}
	\bibfield{author}{\bibinfo{person}{ray project}.}
	\bibinfo{year}{2022}\natexlab{}.
	\newblock \bibinfo{booktitle}{\emph{RayFed}}.
	\newblock
	\urldef\tempurl%
	\url{https://github.com/ray-project/rayfed}
	\showURL{%
		\tempurl}
	
	
	\bibitem[Reina et~al\mbox{.}(2021)]%
	{reina2021openfl}
	\bibfield{author}{\bibinfo{person}{G.~Anthony Reina}, \bibinfo{person}{Alexey
			Gruzdev}, \bibinfo{person}{Patrick Foley}, \bibinfo{person}{Olga
			Perepelkina}, \bibinfo{person}{Mansi Sharma}, \bibinfo{person}{Igor
			Davidyuk}, \bibinfo{person}{Ilya Trushkin}, \bibinfo{person}{Maksim
			Radionov}, \bibinfo{person}{Aleksandr Mokrov}, \bibinfo{person}{Dmitry
			Agapov}, \bibinfo{person}{Jason Martin}, \bibinfo{person}{Brandon Edwards},
		\bibinfo{person}{Micah~J. Sheller}, \bibinfo{person}{Sarthak Pati},
		\bibinfo{person}{Prakash~Narayana Moorthy}, \bibinfo{person}{Hans~Shih{-}Han
			Wang}, \bibinfo{person}{Prashant Shah}, {and} \bibinfo{person}{Spyridon
			Bakas}.} \bibinfo{year}{2021}\natexlab{}.
	\newblock \showarticletitle{OpenFL: An open-source framework for Federated
		Learning}.
	\newblock \bibinfo{journal}{\emph{CoRR}}  \bibinfo{volume}{abs/2105.06413}
	(\bibinfo{year}{2021}).
	\newblock
	
	
	\bibitem[Research(2021)]%
	{FLSim}
	\bibfield{author}{\bibinfo{person}{Meta Research}.}
	\bibinfo{year}{2021}\natexlab{}.
	\newblock \bibinfo{booktitle}{\emph{{Federated Learning Simulator (FLSim)}}}.
	\newblock
	\urldef\tempurl%
	\url{https://github.com/facebookresearch/FLSim}
	\showURL{%
		\tempurl}
	
	
	\bibitem[Ro et~al\mbox{.}(2021)]%
	{FedJAX}
	\bibfield{author}{\bibinfo{person}{Jae~Hun Ro},
		\bibinfo{person}{Ananda~Theertha Suresh}, {and} \bibinfo{person}{Ke Wu}.}
	\bibinfo{year}{2021}\natexlab{}.
	\newblock \showarticletitle{FedJAX: Federated learning simulation with {JAX}}.
	\newblock \bibinfo{journal}{\emph{CoRR}}  \bibinfo{volume}{abs/2108.02117}
	(\bibinfo{year}{2021}).
	\newblock
	
	
	\bibitem[Roth et~al\mbox{.}(2023)]%
	{RothCWY0HKHZL0023}
	\bibfield{author}{\bibinfo{person}{Holger~R. Roth}, \bibinfo{person}{Yan
			Cheng}, \bibinfo{person}{Yuhong Wen}, \bibinfo{person}{Isaac Yang},
		\bibinfo{person}{Ziyue Xu}, \bibinfo{person}{Yuan{-}Ting Hsieh},
		\bibinfo{person}{Kristopher Kersten}, \bibinfo{person}{Ahmed Harouni},
		\bibinfo{person}{Can Zhao}, \bibinfo{person}{Kevin Lu},
		\bibinfo{person}{Zhihong Zhang}, \bibinfo{person}{Wenqi Li},
		\bibinfo{person}{Andriy Myronenko}, \bibinfo{person}{Dong Yang},
		\bibinfo{person}{Sean Yang}, \bibinfo{person}{Nicola Rieke},
		\bibinfo{person}{Abood Quraini}, \bibinfo{person}{Chester Chen},
		\bibinfo{person}{Daguang Xu}, \bibinfo{person}{Nic Ma},
		\bibinfo{person}{Prerna Dogra}, \bibinfo{person}{Mona Flores}, {and}
		\bibinfo{person}{Andrew Feng}.} \bibinfo{year}{2023}\natexlab{}.
	\newblock \showarticletitle{{NVIDIA} {FLARE:} Federated Learning from
		Simulation to Real-World}.
	\newblock \bibinfo{journal}{\emph{{IEEE} Data Eng. Bull.}}
	\bibinfo{volume}{46}, \bibinfo{number}{1} (\bibinfo{year}{2023}),
	\bibinfo{pages}{170--184}.
	\newblock
	
	
	\bibitem[Ryffel et~al\mbox{.}(2018)]%
	{RyffelTDWMRP18}
	\bibfield{author}{\bibinfo{person}{Th{\'{e}}o Ryffel}, \bibinfo{person}{Andrew
			Trask}, \bibinfo{person}{Morten Dahl}, \bibinfo{person}{Bobby Wagner},
		\bibinfo{person}{Jason Mancuso}, \bibinfo{person}{Daniel Rueckert}, {and}
		\bibinfo{person}{Jonathan Passerat{-}Palmbach}.}
	\bibinfo{year}{2018}\natexlab{}.
	\newblock \showarticletitle{A generic framework for privacy preserving deep
		learning}.
	\newblock \bibinfo{journal}{\emph{CoRR}}  \bibinfo{volume}{abs/1811.04017}
	(\bibinfo{year}{2018}).
	\newblock
	
	
	\bibitem[Ryu et~al\mbox{.}(2022)]%
	{appfl-ipdps22}
	\bibfield{author}{\bibinfo{person}{Minseok Ryu}, \bibinfo{person}{Youngdae
			Kim}, \bibinfo{person}{Kibaek Kim}, {and} \bibinfo{person}{Ravi~K. Madduri}.}
	\bibinfo{year}{2022}\natexlab{}.
	\newblock \showarticletitle{{APPFL:} Open-Source Software Framework for
		Privacy-Preserving Federated Learning}. In \bibinfo{booktitle}{\emph{{IEEE}
			International Parallel and Distributed Processing Symposium, {IPDPS}
			Workshops 2022}}.
	\newblock
	
	
	\bibitem[Seaman(1999)]%
	{Seaman99}
	\bibfield{author}{\bibinfo{person}{Carolyn~B. Seaman}.}
	\bibinfo{year}{1999}\natexlab{}.
	\newblock \showarticletitle{Qualitative Methods in Empirical Studies of
		Software Engineering}.
	\newblock \bibinfo{journal}{\emph{{IEEE} Trans. Software Eng.}}
	\bibinfo{volume}{25}, \bibinfo{number}{4} (\bibinfo{year}{1999}),
	\bibinfo{pages}{557--572}.
	\newblock
	
	
	\bibitem[SecretFlow(2022)]%
	{SecretFlow}
	\bibfield{author}{\bibinfo{person}{SecretFlow}.}
	\bibinfo{year}{2022}\natexlab{}.
	\newblock \bibinfo{booktitle}{\emph{{SecretFlow}}}.
	\newblock
	\urldef\tempurl%
	\url{https://github.com/secretflow/secretflow}
	\showURL{%
		\tempurl}
	
	
	\bibitem[Shen et~al\mbox{.}(2021)]%
	{ShenM0TCC21}
	\bibfield{author}{\bibinfo{person}{Qingchao Shen}, \bibinfo{person}{Haoyang
			Ma}, \bibinfo{person}{Junjie Chen}, \bibinfo{person}{Yongqiang Tian},
		\bibinfo{person}{Shing{-}Chi Cheung}, {and} \bibinfo{person}{Xiang Chen}.}
	\bibinfo{year}{2021}\natexlab{}.
	\newblock \showarticletitle{A comprehensive study of deep learning compiler
		bugs}. In \bibinfo{booktitle}{\emph{Proceedings of the 29th {ACM} Joint
			European Software Engineering Conference and Symposium on the Foundations of
			Software Engineering ({ESEC/FSE})}}.
	\newblock
	
	
	\bibitem[Su and Li(2022)]%
	{plato-iwqos22}
	\bibfield{author}{\bibinfo{person}{Ningxin Su} {and} \bibinfo{person}{Baochun
			Li}.} \bibinfo{year}{2022}\natexlab{}.
	\newblock \showarticletitle{How Asynchronous can Federated Learning Be?}. In
	\bibinfo{booktitle}{\emph{30th {IEEE/ACM} International Symposium on Quality
			of Service, IWQoS 2022}}.
	\newblock
	
	
	\bibitem[Sun et~al\mbox{.}(2021)]%
	{SunZWDBC21}
	\bibfield{author}{\bibinfo{person}{Xiaobing Sun}, \bibinfo{person}{Tianchi
			Zhou}, \bibinfo{person}{Rongcun Wang}, \bibinfo{person}{Yucong Duan},
		\bibinfo{person}{Lili Bo}, {and} \bibinfo{person}{Jianming Chang}.}
	\bibinfo{year}{2021}\natexlab{}.
	\newblock \showarticletitle{Experience report: investigating bug fixes in
		machine learning frameworks/libraries}.
	\newblock \bibinfo{journal}{\emph{Frontiers Comput. Sci.}}
	\bibinfo{volume}{15}, \bibinfo{number}{6} (\bibinfo{year}{2021}),
	\bibinfo{pages}{156212}.
	\newblock
	
	
	\bibitem[Tan et~al\mbox{.}(2014)]%
	{TanLLWZZ14}
	\bibfield{author}{\bibinfo{person}{Lin Tan}, \bibinfo{person}{Chen Liu},
		\bibinfo{person}{Zhenmin Li}, \bibinfo{person}{Xuanhui Wang},
		\bibinfo{person}{Yuanyuan Zhou}, {and} \bibinfo{person}{ChengXiang Zhai}.}
	\bibinfo{year}{2014}\natexlab{}.
	\newblock \showarticletitle{Bug characteristics in open source software}.
	\newblock \bibinfo{journal}{\emph{Empir. Softw. Eng.}} \bibinfo{volume}{19},
	\bibinfo{number}{6} (\bibinfo{year}{2014}), \bibinfo{pages}{1665--1705}.
	\newblock
	\urldef\tempurl%
	\url{https://doi.org/10.1007/s10664-013-9258-8}
	\showDOI{\tempurl}
	
	
	\bibitem[Thung et~al\mbox{.}(2012)]%
	{ThungWLJ12}
	\bibfield{author}{\bibinfo{person}{Ferdian Thung}, \bibinfo{person}{Shaowei
			Wang}, \bibinfo{person}{David Lo}, {and} \bibinfo{person}{Lingxiao Jiang}.}
	\bibinfo{year}{2012}\natexlab{}.
	\newblock \showarticletitle{An Empirical Study of Bugs in Machine Learning
		Systems}. In \bibinfo{booktitle}{\emph{23rd {IEEE} International Symposium on
			Software Reliability Engineering, {ISSRE} 2012}}.
	\newblock
	
	
	\bibitem[Tizpaz{-}Niari et~al\mbox{.}(2020)]%
	{NiariC020}
	\bibfield{author}{\bibinfo{person}{Saeid Tizpaz{-}Niari},
		\bibinfo{person}{Pavol Cern{\'{y}}}, {and} \bibinfo{person}{Ashutosh
			Trivedi}.} \bibinfo{year}{2020}\natexlab{}.
	\newblock \showarticletitle{Detecting and understanding real-world differential
		performance bugs in machine learning libraries}. In
	\bibinfo{booktitle}{\emph{Proceedings of the 29th {ACM} {SIGSOFT}
			International Symposium on Software Testing and Analysis ({ISSTA})}}.
	\bibinfo{pages}{189--199}.
	\newblock
	
	
	\bibitem[Wen et~al\mbox{.}(2018)]%
	{WenMing2018}
	\bibfield{author}{\bibinfo{person}{Ming Wen}, \bibinfo{person}{Junjie Chen},
		\bibinfo{person}{Rongxin Wu}, \bibinfo{person}{Dan Hao}, {and}
		\bibinfo{person}{Shing-Chi Cheung}.} \bibinfo{year}{2018}\natexlab{}.
	\newblock \showarticletitle{Context-Aware Patch Generation for Better Automated
		Program Repair}. In \bibinfo{booktitle}{\emph{Proceedings of the 40th
			International Conference on Software Engineering}}.
	\newblock
	
	
	\bibitem[Weng et~al\mbox{.}(2021)]%
	{WengWZLZL21}
	\bibfield{author}{\bibinfo{person}{Jia{-}Si Weng}, \bibinfo{person}{Jian Weng},
		\bibinfo{person}{Jilian Zhang}, \bibinfo{person}{Ming Li},
		\bibinfo{person}{Yue Zhang}, {and} \bibinfo{person}{Weiqi Luo}.}
	\bibinfo{year}{2021}\natexlab{}.
	\newblock \showarticletitle{DeepChain: Auditable and Privacy-Preserving Deep
		Learning with Blockchain-Based Incentive}.
	\newblock \bibinfo{journal}{\emph{{IEEE} Trans. Dependable Secur. Comput.}}
	\bibinfo{volume}{18}, \bibinfo{number}{5} (\bibinfo{year}{2021}),
	\bibinfo{pages}{2438--2455}.
	\newblock
	
	
	\bibitem[Williams and Cockburn(2003)]%
	{WilliamsC03}
	\bibfield{author}{\bibinfo{person}{Laurie~A. Williams} {and}
		\bibinfo{person}{Alistair Cockburn}.} \bibinfo{year}{2003}\natexlab{}.
	\newblock \showarticletitle{Guest Editors' Introduction: Agile Software
		Development: It's about Feedback and Change}.
	\newblock \bibinfo{journal}{\emph{Computer}} \bibinfo{volume}{36},
	\bibinfo{number}{6} (\bibinfo{year}{2003}), \bibinfo{pages}{39--43}.
	\newblock
	
	
	\bibitem[Wu et~al\mbox{.}(2022)]%
	{FedSim}
	\bibfield{author}{\bibinfo{person}{Zhaomin Wu}, \bibinfo{person}{Qinbin Li},
		{and} \bibinfo{person}{Bingsheng He}.} \bibinfo{year}{2022}\natexlab{}.
	\newblock \showarticletitle{A Coupled Design of Exploiting Record Similarity
		for Practical Vertical Federated Learning}. In
	\bibinfo{booktitle}{\emph{NeurIPS}}.
	\newblock
	
	
	\bibitem[Xia and Zhang(2022)]%
	{Xia2022LessTM}
	\bibfield{author}{\bibinfo{person}{Chun Xia} {and} \bibinfo{person}{Lingming
			Zhang}.} \bibinfo{year}{2022}\natexlab{}.
	\newblock \showarticletitle{Less training, more repairing please: revisiting
		automated program repair via zero-shot learning}.
	\newblock \bibinfo{journal}{\emph{Proceedings of the 30th ACM Joint European
			Software Engineering Conference and Symposium on the Foundations of Software
			Engineering}} (\bibinfo{year}{2022}).
	\newblock
	
	
	\bibitem[Xie et~al\mbox{.}(2023)]%
	{federatedscope}
	\bibfield{author}{\bibinfo{person}{Yuexiang Xie}, \bibinfo{person}{Zhen Wang},
		\bibinfo{person}{Dawei Gao}, \bibinfo{person}{Daoyuan Chen},
		\bibinfo{person}{Liuyi Yao}, \bibinfo{person}{Weirui Kuang},
		\bibinfo{person}{Yaliang Li}, \bibinfo{person}{Bolin Ding}, {and}
		\bibinfo{person}{Jingren Zhou}.} \bibinfo{year}{2023}\natexlab{}.
	\newblock \showarticletitle{FederatedScope: {A} Flexible Federated Learning
		Platform for Heterogeneity}.
	\newblock \bibinfo{journal}{\emph{Proc. {VLDB} Endow.}} \bibinfo{volume}{16},
	\bibinfo{number}{5} (\bibinfo{year}{2023}), \bibinfo{pages}{1059--1072}.
	\newblock
	
	
	\bibitem[Xu et~al\mbox{.}(2020)]%
	{XuLL0L20}
	\bibfield{author}{\bibinfo{person}{Guowen Xu}, \bibinfo{person}{Hongwei Li},
		\bibinfo{person}{Sen Liu}, \bibinfo{person}{Kan Yang}, {and}
		\bibinfo{person}{Xiaodong Lin}.} \bibinfo{year}{2020}\natexlab{}.
	\newblock \showarticletitle{VerifyNet: Secure and Verifiable Federated
		Learning}.
	\newblock \bibinfo{journal}{\emph{{IEEE} Trans. Inf. Forensics Secur.}}
	\bibinfo{volume}{15} (\bibinfo{year}{2020}), \bibinfo{pages}{911--926}.
	\newblock
	
	
	\bibitem[Yang et~al\mbox{.}(2019)]%
	{YangLCT19}
	\bibfield{author}{\bibinfo{person}{Qiang Yang}, \bibinfo{person}{Yang Liu},
		\bibinfo{person}{Tianjian Chen}, {and} \bibinfo{person}{Yongxin Tong}.}
	\bibinfo{year}{2019}\natexlab{}.
	\newblock \showarticletitle{Federated Machine Learning: Concept and
		Applications}.
	\newblock \bibinfo{journal}{\emph{{ACM} Trans. Intell. Syst. Technol.}}
	\bibinfo{volume}{10}, \bibinfo{number}{2} (\bibinfo{year}{2019}),
	\bibinfo{pages}{12:1--12:19}.
	\newblock
	
	
	\bibitem[Yang et~al\mbox{.}(2022)]%
	{YangHXF22}
	\bibfield{author}{\bibinfo{person}{Yilin Yang}, \bibinfo{person}{Tianxing He},
		\bibinfo{person}{Zhilong Xia}, {and} \bibinfo{person}{Yang Feng}.}
	\bibinfo{year}{2022}\natexlab{}.
	\newblock \showarticletitle{A comprehensive empirical study on bug
		characteristics of deep learning frameworks}.
	\newblock \bibinfo{journal}{\emph{Inf. Softw. Technol.}}  \bibinfo{volume}{151}
	(\bibinfo{year}{2022}), \bibinfo{pages}{107004}.
	\newblock
	
	
	\bibitem[Yin et~al\mbox{.}(2022)]%
	{YinZH21}
	\bibfield{author}{\bibinfo{person}{Xuefei Yin}, \bibinfo{person}{Yanming Zhu},
		{and} \bibinfo{person}{Jiankun Hu}.} \bibinfo{year}{2022}\natexlab{}.
	\newblock \showarticletitle{A Comprehensive Survey of Privacy-preserving
		Federated Learning: {A} Taxonomy, Review, and Future Directions}.
	\newblock \bibinfo{journal}{\emph{{ACM} Comput. Surv.}} \bibinfo{volume}{54},
	\bibinfo{number}{6} (\bibinfo{year}{2022}), \bibinfo{pages}{131:1--131:36}.
	\newblock
	
	
	\bibitem[Zeng et~al\mbox{.}(2023)]%
	{FedLab}
	\bibfield{author}{\bibinfo{person}{Dun Zeng}, \bibinfo{person}{Siqi Liang},
		\bibinfo{person}{Xiangjing Hu}, \bibinfo{person}{Hui Wang}, {and}
		\bibinfo{person}{Zenglin Xu}.} \bibinfo{year}{2023}\natexlab{}.
	\newblock \showarticletitle{FedLab: A Flexible Federated Learning Framework}.
	\newblock \bibinfo{journal}{\emph{Journal of Machine Learning Research}}
	\bibinfo{volume}{24}, \bibinfo{number}{100} (\bibinfo{year}{2023}),
	\bibinfo{pages}{1--7}.
	\newblock
	
	
	\bibitem[Zhang et~al\mbox{.}(2020)]%
	{ZhangLXWYL20}
	\bibfield{author}{\bibinfo{person}{Chengliang Zhang}, \bibinfo{person}{Suyi
			Li}, \bibinfo{person}{Junzhe Xia}, \bibinfo{person}{Wei Wang},
		\bibinfo{person}{Feng Yan}, {and} \bibinfo{person}{Yang Liu}.}
	\bibinfo{year}{2020}\natexlab{}.
	\newblock \showarticletitle{{BatchCrypt}: Efficient Homomorphic Encryption for
		{Cross-Silo} Federated Learning}. In \bibinfo{booktitle}{\emph{2020 {USENIX}
			Annual Technical Conference, {USENIX} {ATC} 2020}}.
	\newblock
	
	
	\bibitem[Zhang et~al\mbox{.}(2019)]%
	{ZhangGMLK19}
	\bibfield{author}{\bibinfo{person}{Tianyi Zhang}, \bibinfo{person}{Cuiyun Gao},
		\bibinfo{person}{Lei Ma}, \bibinfo{person}{Michael~R. Lyu}, {and}
		\bibinfo{person}{Miryung Kim}.} \bibinfo{year}{2019}\natexlab{}.
	\newblock \showarticletitle{An Empirical Study of Common Challenges in
		Developing Deep Learning Applications}. In \bibinfo{booktitle}{\emph{30th
			{IEEE} International Symposium on Software Reliability Engineering, {ISSRE}
			2019}}.
	\newblock
	
	
	\bibitem[Zhang et~al\mbox{.}(2018)]%
	{ZhangCCXZ18}
	\bibfield{author}{\bibinfo{person}{Yuhao Zhang}, \bibinfo{person}{Yifan Chen},
		\bibinfo{person}{Shing{-}Chi Cheung}, \bibinfo{person}{Yingfei Xiong}, {and}
		\bibinfo{person}{Lu Zhang}.} \bibinfo{year}{2018}\natexlab{}.
	\newblock \showarticletitle{An empirical study on TensorFlow program bugs}. In
	\bibinfo{booktitle}{\emph{Proceedings of the 27th {ACM} {SIGSOFT}
			International Symposium on Software Testing and Analysis ({ISSTA})}}.
	\bibinfo{pages}{129--140}.
	\newblock
	
	
	\bibitem[Zhou et~al\mbox{.}(2021)]%
	{abs-2104-10501}
	\bibfield{author}{\bibinfo{person}{Jiehan Zhou}, \bibinfo{person}{Shouhua
			Zhang}, \bibinfo{person}{Qinghua Lu}, \bibinfo{person}{Wenbin Dai},
		\bibinfo{person}{Min Chen}, \bibinfo{person}{Xin Liu},
		\bibinfo{person}{Susanna Pirttikangas}, \bibinfo{person}{Yang Shi},
		\bibinfo{person}{Weishan Zhang}, {and} \bibinfo{person}{Enrique
			Herrera{-}Viedma}.} \bibinfo{year}{2021}\natexlab{}.
	\newblock \showarticletitle{A Survey on Federated Learning and its Applications
		for Accelerating Industrial Internet of Things}.
	\newblock \bibinfo{journal}{\emph{CoRR}}  \bibinfo{volume}{abs/2104.10501}
	(\bibinfo{year}{2021}).
	\newblock
	
	
\end{thebibliography}


\end{document}